\documentclass[useAMS,usenatbib,epsfig]{mn2e}
\newif\ifAMStwofonts
\AMStwofontstrue
\usepackage{amsmath,amsfonts,epsfig,natbib}

\def\reff@jnl#1{{\rm#1\/}}
\def\aj{\reff@jnl{AJ}}                  % Astronomical Journal
\def\araa{\reff@jnl{ARA\&A}}            % Annual Review of Astron and Astrophys
\def\apj{\reff@jnl{ApJ}}                % Astrophysical Journal
\def\apjl{\reff@jnl{ApJ}}               % Astrophysical Journal, Letters
\def\apjs{\reff@jnl{ApJS}}              % Astrophysical Journal, Supplement
\def\ao{\reff@jnl{Appl.Optics}}         % Applied Optics
\def\apss{\reff@jnl{Ap\&SS}}            % Astrophysics and Space Science
\def\aap{\reff@jnl{A\&A}}               % Astronomy and Astrophysics
\def\aapr{\reff@jnl{A\&A~Rev.}}         % Astronomy and Astrophysics Reviews
\def\aaps{\reff@jnl{A\&AS}}             % Astronomy and Astrophysics, Supplement
\def\azh{\reff@jnl{AZh}}                % Astronomicheskii Zhurnal
\def\baas{\reff@jnl{BAAS}}              % Bulletin of the AAS
\def\gca{\reff@jnl{GeCoA}}              % Geochimica et Cosmochimica Acta
\def\jrasc{\reff@jnl{JRASC}}            % Journal of the RAS of Canada
\def\memras{\reff@jnl{MmRAS}}           % Memoirs of the RAS
\def\mnras{\reff@jnl{MNRAS}}            % Monthly Notices of the RAS
\def\pra{\reff@jnl{Phys.Rev.A}}         % Physical Review A: General Physics
\def\prb{\reff@jnl{Phys.Rev.B}}         % Physical Review B: Solid State
\def\prc{\reff@jnl{Phys.Rev.C}}         % Physical Review C
\def\prd{\reff@jnl{Phys.Rev.D}}         % Physical Review D
\def\prl{\reff@jnl{Phys.Rev.Lett}}      % Physical Review Letters
\def\pasp{\reff@jnl{PASP}}              % Publications of the ASP
\def\pasj{\reff@jnl{PASJ}}              % Publications of the ASJ
\def\qjras{\reff@jnl{QJRAS}}            % Quarterly Journal of the RAS
\def\skytel{\reff@jnl{S\&T}}            % Sky and Telescope
\def\solphys{\reff@jnl{Solar~Phys.}}    % Solar Physics
\def\sovast{\reff@jnl{Soviet~Ast.}}     % Soviet Astronomy
\def\ssr{\reff@jnl{Space~Sci.Rev.}}     % Space Science Reviews
\def\zap{\reff@jnl{ZAp}}                % Zeitschrift fuer Astrophysik
\def\nat{\reff@jnl{Nature}}             % Nature

\newcommand{\wmap}{{\it WMAP }}
\newcommand{\planck}{{\it Planck }}
\def\lesssim{\mathrel{\hbox{\rlap{\hbox{\lower4pt\hbox{$\sim$}}}\hbox{$<$}}}}

\title[AME observations with QUIJOTE]
{Measurements of the Intensity and Polarization of the Anomalous Microwave Emission in the Perseus molecular complex with QUIJOTE}

\author[R. G\'enova-Santos et al.] {R. G\'enova-Santos$^{1,5}\thanks{E-mail: rgs@iac.es}$, J.~A. Rubi\~no Mart\'{\i}n$^{1,5}$, R. Rebolo$^{1,5,6}$,  A. Pel\'aez-Santos$^{1,5}$,
\newauthor C.~H. L\'opez-Caraballo$^{1,5,7}$, S. Harper$^{2}$, R.~A. Watson$^{2}$, M.~Ashdown$^{4,8}$, 
\newauthor R.~B. Barreiro$^{3}$, B. Casaponsa$^{3}$, C. Dickinson$^{2}$,  J.~M. Diego$^{3}$, R. Fern\'andez-Cobos$^{3}$, 
\newauthor K.~J.~B. Grainge$^{4,2}$, D. Herranz$^{3}$, R. Hoyland$^1$, A. Lasenby$^{4,8}$, M. L\'opez-Caniego$^{3}$, 
\newauthor E. Mart\'{\i}nez-Gonz\'alez$^{3}$, M. McCulloch$^{2}$, S. Melhuish$^{2}$, L. Piccirillo$^{2}$, Y.~C. Perrott$^4$, 
\newauthor F. Poidevin$^{1,5}$, N. Razavi-Ghods$^{4}$, P.F. Scott$^{4}$, D. Titterington$^{4}$, D. Tramonte$^{1,5}$, 
\newauthor P. Vielva$^{3}$ and R. Vignaga$^{1,5}$\\
$^1$ Instituto de Astrofis\'{i}ca de Canarias, 38200 La Laguna, Tenerife, Canary Islands, Spain\\
$^2$ Jodrell Bank Centre for Astrophysics, Alan Turing Building, School of Physics and Astronomy, The University of Manchester,\\
Oxford Road, Manchester, M13 9PL, U.K\\ 
$^3$ Instituto de F\'{\i}sica de Cantabria (CSIC-Universidad de Cantabria), Avda. de los Castros s/n, 39005 Santander, Spain\\
$^4$ Astrophysics Group, Cavendish Laboratory, University of Cambridge, J.J. Thomson Avenue, Cambridge CB3 0HE, UK\\
$^5$ Departamento de Astrof\'{\i}sica, Universidad de La Laguna (ULL), 38206 La Laguna, Tenerife, Spain\\
$^6$ Consejo Superior de Investigaciones Cient\'{\i}ficas, Spain\\
$^7$ Departamento de F\'{\i}sica, Universidad de la Serena, Av. Cisternas 1200, La Serena, Chile\\
$^8$ Kavli Institute for Cosmology, Madingley Road, Cambridge, CB3 0HA
}
\date{Accepted Received In original form}

\pagerange{\pageref{firstpage}--\pageref{lastpage}}
\pubyear{2014}
\begin{document}

\label{firstpage}
\maketitle

\begin{abstract}
Anomalous microwave emission (AME) has been observed in numerous sky regions, using different experiments in the frequency 
range $\sim 10-60$~GHz. One of the most scrutinized regions is G159.6-18.5, located within the Perseus molecular complex. 
In this paper we present further observations of this region (194~hours in total over $\approx 250$~deg$^2$), both 
in intensity and in polarization.  They span four independent frequency channels between 10 and 20~GHz, and were gathered 
with QUIJOTE, a new CMB experiment with the goal of measuring the polarization of the CMB and Galactic foregrounds. When 
combined with other publicly-available intensity data, we achieve the most precise spectrum of the AME measured to date in 
an individual region, with 13 independent data points between 10 and 50~GHz being dominated by this emission. 
The four QUIJOTE data points provide the first independent confirmation of the downturn of the AME spectrum at low frequencies, 
initially unveiled by the COSMOSOMAS experiment in this region. 
We accomplish an accurate fit of these data using models based on electric dipole emission 
from spinning dust grains, and also fit some of the parameters on which these models depend.

We also present polarization maps with an angular resolution of $\approx 1^\circ$ and a sensitivity of $\approx 25~\mu$K/beam. 
From these maps, which are consistent with zero polarization, we obtain upper limits of $\Pi<6.3\%$ and $<2.8\%$ ($95\%$ C.L.) 
respectively at 12 and 18~GHz, a frequency range where no AME polarization observations have been reported to date.
These constraints are compatible with theoretical predictions of the polarization fraction from electric dipole emission 
originating from spinning dust grains. 
At the same time, they rule out several models based on magnetic dipole emission from dust grains ordered in a single 
magnetic domain, which typically predict higher polarization levels. Future QUIJOTE data in this region may allow more 
stringent constraints on the polarization level of the AME.
\end{abstract}

\begin{keywords}
cosmology: cosmic microwave background -  radiation mechanisms: general - ISM: individual objects: G159.6-18.5 - diffuse radiation - 
radio continuum: ISM.
\end{keywords}

\section{Introduction}

Now that the observations of the temperature anisotropies of the Cosmic Microwave Background (CMB) have reached levels of 
sensitivity and angular resolution that have allowed the determination of the main cosmological parameters with accuracies 
close to $1\%$ \citep{bennett13,cpp1}, attention has shifted to the study of the polarization of these anisotropies. 
CMB polarization also encodes a wealth of cosmological information. In particular, it is of paramount importance to 
study the B-mode anisotropy, which can only be created by vector or tensor perturbations, such as those due to primordial 
gravitational waves \citep{kamionkowsky97,zaldarriaga97}. As all inflationary scenarios are predicted to give rise to 
gravitational waves, this B-mode signal is a definite test for inflation. Following previous upper limits placed by experiments 
like QUIET \citep{quiet12} or BICEP \citep{bicep114}, the first detection of this signal was claimed in data from the 
BICEP2 experiment at 150~GHz, with a level of the tensor-to-scalar ratio $r=0.20^{+0.07}_{-0.05}$ \citep{bicep214}. 
However, soon after these results were published, some papers \citep{flauger14,mortonson14} appeared 
pointing out that the BICEP2 team could have underestimated the contribution from polarized thermal dust emission, 
leading to a wrong interpretation of their signal. This has been recently supported by data from the \planck satellite, 
which have shown that the level of polarized dust emission in the region of the sky covered by BICEP2 
could form a significant component of the measured signal \citep{pip30}, thus causing a likely reduction in the level 
of cosmological signal that can be inferred from the BICEP2 results.

It is therefore clear that, any unambiguous B-mode detection requires a full assessment of the level of foreground 
contamination and ideally confirmation by independent experiments operating at different frequencies. QUIJOTE is one of 
these experiments, which operates in the frequency range 10-40~GHz, and is subject to different systematics and foregrounds. 
QUIJOTE data will also be useful for measuring the polarization of low-frequency 
foregrounds, which are known to dominate the underlying primordial B-mode polarization over a wide frequency range, and 
therefore need to be characterized and removed from the observed maps. Synchrotron emission from relativistic electrons 
is the most important contaminant at low frequencies in polarization. In intensity, free-free emission and the so-called 
anomalous microwave emission (AME) also show up in this frequency range. While the former is known to be practically unpolarized \citep{rybicki79}, 
very little is known about the polarization level of the AME. Since its discovery in the 90s \citep{kogut96a,kogut96b,leitch97}, 
many observations in large-sky areas \citep{oliveira98,oliveira99,davies06} and in individual Galactic
\citep{finkbeiner02,watson05,casassus06,dickinson09,ami09,tibbs10,genova11,vidal11,per20,pip25} and extragalactic \citep{murphy10} clouds 
have contributed to the understanding of the physical properties of this emission. A great deal of effort has also been dedicated 
to theoretical modelling of AME. Electric dipole radiation from very small and rapidly rotating dust grains in the interstellar medium 
\citep{draine98,ali09,hoang10,ysard10,silsbee11}, the so-called {\it spinning dust emission}, is the scenario that best fits the 
observations. An alternative mechanism based on magnetic dipole emission has also been proposed \citep{draine99}, with 
a spectrum peaking at higher frequencies.

The characterization of AME polarization properties is important not only to get further insight into its physical mechanism, but 
to assess how it could affect experiments pursuing the primordial B-mode signal in the frequency range 10 to 50 GHz. There are some 
theoretical studies of the AME polarization in the literature. After the first predictions by \citet{lazarian00} of 
spinning dust polarization, more recently \citet{hoang13} presented a model based on observations of the UV polarization bump,
according to which the maximum polarization fraction would be $\approx 3\%$, at a frequency of 5~GHz. In regard to the magnetic 
dipole emission model, while \citet{draine99} predicted high polarization fractions (up to 40~\%) in the case of dust 
grains with atomic magnetic moments oriented in a single domain, \citet{draine13} recently presented a more realistic model, with 
randomly-oriented magnetic inclusions, which results in lower polarization degrees ($\la 5$\% in the range $10-20$~GHz). However, not 
much is known, from the observational standpoint, about the polarization properties of the AME. Using \wmap 5-year data 
\citet{macellari11} set an upper limit $\Pi<5\%$ (this and other upper limits that will be referred to in this section are at the 95\% 
confidence level) on the polarization fraction of the diffuse AME. Some other constraints refer to individual clouds. 
\citet{battistelli06} observed the Perseus molecular complex with the COSMOSOMAS experiment and derived $\Pi=3.4^{1.5}_{-1.9}\%$ at 
11~GHz. \citet{casassus08} reported an upper limit of $<1.0\%$ at 31~GHz on the $\rho$ Ophiuchi molecular cloud using the 
{\it Cosmic Background Imager}, whereas \citet{mason09} found a maximum of $<2.7\%$ with the Green Bank Telescope at 9~GHz. More recently, 
\citet{lopez11} obtained an upper limit of $<1.0\%$ at 23~GHz on the Perseus molecular complex using \wmap 7-year data. Shortly after 
this paper \citet{dickinson11}, using the same data, obtained $<1.4\%$ in the same region, and $<1.7\%$ in $\rho$ Ophiuchi. A detailed 
review of all these observations, plus some updated constraints in some regions, have been presented in \citet{rubino12b}.

In this paper we present the first results obtained with the QUIJOTE experiment, that are based on observations of G159.6-18.5 in the 
Perseus molecular complex, one of the most studied AME regions in the sky \citep{watson05,tibbs10,tibbs13,per20,pip25}. QUIJOTE observations 
cover the frequency range $10-20$~GHz, where only the COSMOSOMAS experiment had provided observations of the AME before. 
The goal of this paper is twofold: to confirm the downturn of the AME spectrum at frequencies below 23~GHz through similar spectral 
sampling but completely independent results to those provided by COSMOSOMAS, and to set constraints on the polarization level 
of the AME in the so far unexplored spectral region between 12 and 20~GHz.
Section~2 is dedicated to the description of the observations and the basics of the data reduction. 
Our main results are presented in section~3, while the conclusions of this work are discussed in section~4.

\section{Data and methodology}

\subsection{QUIJOTE data}\label{sec:quijote_data}

The new data presented in this article were acquired with the QUIJOTE experiment. QUIJOTE is a collaborative project that 
consists of two telescopes and three polarimeter instruments covering respectively the frequencies 10-20, 30 and 
40~GHz, and located at the Teide Observatory (2400~m a.s.l.) in Tenerife (Spain).
The main science driver of this experiment is to perform observations of the CMB polarization to constrain the B-mode 
signal down to $r=0.05$, and to characterize the polarization of low-frequency foregrounds, mainly synchrotron emission 
and the AME, so that this signal can be removed from the primordial maps to a level that will permit reaching the previous 
level of $r$. The two QUIJOTE telescopes are based on an offset crossed-Dragone design, with projected apertures of $2.25$~m 
and $1.89$~m for the primary and secondary mirrors, and provide highly symmetric beams (ellipticity $> 0.98$) with very 
low sidelobes ($\leq -40$~dB) and polarization leakage ($\leq -25$~dB). The first instrument to be fielded on the QUIJOTE 
first telescope is a multi-frequency instrument (MFI) with 4 horns covering the frequency range $10-20$~GHz, and with angular 
resolutions close to one degree. These detectors 
are fitted with MMIC low-noise amplifiers (noise temperature better than 10~K), and use stepped polar modulators to measure 
the polarization of the incoming radiation, providing instantaneous sensitivities of $\approx 650~\mu$K$~s^{1/2}$ in four individual 
frequencies with nominal figures: 11, 13, 17 and 19~GHz. The median integrated PWV above the observatory is $\approx 4$~mm, giving a zenith atmosphere 
temperature of $\approx 2$~K at 11~GHz and $\approx 5$~K at 19~GHz. The MFI saw first light on November 
2012 and since then it is performing routine observations of different Galactic and cosmological regions. The second instrument 
consists of 31 polarimeters at 30~GHz (TGI, thirty-gigahertz instrument), and is based on the same design of the MFI except that 
the polarization modulation is achieved electronically through phase switches. This instrument will be commissioned during 2015. 
Finally, the third instrument is planned to have 31 polarimeters at 40~GHz (FGI, forty-gigahertz instrument). Using the TGI and the 
FGI, which will provide instantaneous sensitivities of $50~\mu$K~s$^{1/2}$, we plan to survey an area of $3000$~deg$^2$ down to a 
projected sensitivity of $\leq 1~\mu$K/beam. A more detailed 
description of the technical and scientific aspects of this project can be found in \citet{rubino12} or in Rebolo et al. 
(in preparation).

\subsubsection{Observations}

The observations covering the Perseus molecular complex were carried out between December 2012 and April 2013 using the MFI, 
in four frequency bands centred at $11.2$, $12.9$, $16.7$ and 
$18.7$~GHz. The beam FWHMs are $0.87^\circ$, for the two lower frequencies, and $0.65^\circ$, for the two high 
frequency bands. The observations consisted of raster scans at constant elevation in order to minimize the effect of atmospheric variations. 
Each scan had an amplitude in azimuth direction of 12 degrees centred around the coordinates RA = 3$^{\rm h}$52$^{\rm m}$, 
Dec. = 34$^\circ$. This position was chosen to be equidistant between the AME cloud G159.6-18.5 and the California HII region 
(NGC1499), which is also observed in the scans and is used as a null test for zero polarization. In each raster scan, of total 
integration time of 30-35 minutes, the telescope moves back and forth in azimuth at a velocity of 1~deg/s. A total of 336 raster scans 
were performed, in four positions of the polar modulators ($0^\circ$, $22.5^\circ$, $45^\circ$ and $67.5^\circ$) in 
order to minimize systematics. In total 194.4~hours of data were accumulated, 23\% of which were removed due to being affected 
by bad weather or instrumental effects, resulting in a final effective observing time of 148.9~hours. As the MFI horns are 
separated typically by 5~degrees on the sky, the sky coverage of each one is different. The total sky area covered by the four 
horns was respectively 176, 184, 277 and 261~deg$^2$.

\subsubsection{Amplitude calibration}

We determine the gain calibration factors (giving the conversion from voltage measured in the detectors to temperature 
on the sky) using intensity measurements on Cas A, which is ideal as it is bright and has a very low degree of polarization. 
We do daily 25-min raster scans of $10\times 10$ degrees around this source from which we derive the gain calibration factors for 
each channel. The output signal of each channel is a 
combination of the three Stokes parameters $I$, $Q$ and $U$. According to \wmap the polarized intensity of Cas A at 22.8~GHz 
is $P=0.81\pm 0.05$~Jy (polarization fraction $\Pi=0.35\pm 0.02\%$) \citep{weiland11}, which is low enough not to be detected in a 
single raster scan. We can therefore safely assume $Q=U=0$, and use the $I$ flux to calibrate. To achieve this we use the modelled 
spectrum of \citet{weiland11}, which is obtained by fitting a combination of \wmap 7-year data and other ancillary data to a 
logarithmic quadratic function. This function is then integrated over the measured spectral transmission of each frequency band to 
obtain the reference Cas A fluxes associated with each channel. 

Finally, as the previous model is referred to epoch 2000, in order to account for the secular decrease of the Cas A flux 
(typically 0.5\%/year), we use the \citet{hafez2008} model, which was derived using VSA observations, in order to refer 
the final fluxes to the time of the observations. In order to circumvent possible uncertainties associated with this secular 
variation, a more suitable calibration source would be Jupiter. However, owing to its small angular size, this source is severely 
diluted in the QUIJOTE beams and a large number of observations would be required for it to be used as primary calibrator. 
We point out that, in any case, the gain calibration will only affect the modelling of the SED in G159.6-18.5.
The gain calibration factors will cancel out when dividing the polarized intensity by the total intensity, and therefore 
the inferred polarization fractions, which are one of the main goals of this paper, are insensitive to the absolute flux calibration.

\subsubsection{Polarization calibration}\label{sec:polcal}

One of the main steps of the data processing is the calibration of the polarization angle $\varphi_0$, which is defined as the reference 
position angle of each polar modulator. To accomplish this we use Tau A (also known as the Crab Nebula) as a calibrator, which is the brightest 
polarized source in the sky in the microwave range. We perform daily 25-min raster scans of $10\times 10$ degrees around this
source, from which we derive a polarized flux that is a function of the intrinsic $Q/I$ and $U/I$, of the position of the 
modulator relative to $\varphi_0$ and of the parallactic angle $\varphi_{\rm p}$. \wmap 7-year results \citep{weiland11} show that 
the $Q/I$ and $U/I$ ratios for Tau A do not significantly vary (less than 2\%, which is consistent with the error 
of the measurement) between 23 and 94 GHz. We therefore assume that these factors will remain equally unchanged down to 10~GHz, and use 
as reference the \wmap measurements at 22.8~GHz, the closest frequency. As we also know $\varphi_{\rm p}$, we can therefore fit 
for $\varphi_0$.

Using 191 raster scans on Tau A throughout a year we have checked that the recovered polarization angle, $\varphi_0$, is stable 
over time. We then combine all these observations to derive a unique value for each horn. The accuracy on the determination of 
this angle is respectively $0.8^\circ$ and $0.4^\circ$ for the two horns that will be used in the polarization analyses that will 
be presented in this paper. In QUIET, an experiment which has similarities with QUIJOTE, a precision of $0.5^\circ$ is achieved by using a 
combination of Tau A observations with a sparse-wire-grid calibrator \citep{quiet12}. We point out however that the accurate 
determination of this angle is important only to derive precise $Q$ and $U$ fluxes. An incorrect angle will result in a mixing of 
flux between $Q$ and $U$, but the polarized intensity $P=\sqrt{Q^2+U^2}$ will remain unchanged. As in this analysis we will get 
constraints on the polarization fraction, the accurate calibration of the polarization angle is unimportant.

\subsubsection{Map making}

The four output channels of each frequency band contain a combination of three Stokes parameters $I$, $Q$, $U$. The sum of these 
channels, after calibration of their individual gains, give $I$, while the subtraction of pairs of channels gives the following 
combination of $Q$ and $U$:
\begin{equation}
 V_{\rm sub} = Q{\rm sin}(4\varphi_{\rm pm}+2\varphi_{\rm p}) + U{\rm cos}(4\varphi_{\rm pm}+2\varphi_{\rm p})~~,
 \label{eq:v_sub}
\end{equation}
where $\varphi_{\rm pm}$ is the position angle of the polar modulator, whose reference position is calibrated following the procedure 
explained in section~\ref{sec:polcal}, and $\varphi_{\rm p}$ is the parallactic angle. Out of the four channels, two are correlated 
and therefore are affected by the same $1/f$ noise, while the two other are not. To reconstruct the polarization signal we then 
only use the two correlated channels, in order to minimize the $1/f$ contribution. The typical knee frequencies of our receivers 
are $f_{\rm k}\sim 10-40$~Hz depending on the channel. However, for the measurement of polarization, the subtraction results in 
much lower values of $f_{\rm k}\sim 0.1-0.2$~Hz. In order to further reduce the $1/f$ noise in the final maps, we apply a filter on the 
time-ordered-data (TOD) by subtracting the median of the data in intervals of 20~s, after binning the data at 50~ms.

Under the assumption that the filtered TODs are dominated by white noise, we consider the noise covariance matrix to be diagonal, 
a hypothesis that considerably simplifies the map-making. As the response of our instrument to polarization is a combination of 
$Q$ and $U$, in order to recover these Stokes parameters in each pixel we have to combine all the samples lying in that pixel 
corresponding to different angles $\varphi=4\varphi_{\rm pm}+2\varphi_{\rm p}$. To do so we use two independent strategies. The first 
one consists on producing 100 maps, each one corresponding to $\varphi$ angles within a given bin, and then using the 100 values of 
each pixel to find the best-fit solution for $Q$ and $U$ from equation~\ref{eq:v_sub}. The second strategy builds on an analytical 
$\chi^2$ minimization. The different parameters, which are combinations of sines and cosines of the $\varphi$ angles, are grouped in 
each pixel, and at the end of the process the $Q$ and $U$ values are computed using the analytical formulas that result from this 
minimization. In both cases the data samples are weighted according to their noise, which is calculated from the standard deviation 
calculated during the binning of the TODs. 

To produce the final maps we use a HEALPix pixelization 
\citep{gorski05} with $N_{\rm side}=512$ (pixel size $6.9$~arcmin), which is sufficient given the beam FWHM. While we have checked that 
the maps resulting from the two strategies are almost identical, in the subsequent analyses we use those resulting from the second 
method.

\subsection{Ancillary data}\label{sec:ancillary}

All the polarization data that will be used in this article comes from the QUIJOTE experiment. However, in order to obtain the 
full spectral energy distribution (SED) of G159.6-18.5, from which the residual AME fluxes will be inferred, we 
use ancillary data from other experiments. In the low-frequency range we use the \citet{haslam82} 
map\footnote{We use the map supplied by \citet{platania03}.} at 0.408~GHz, the \citet{berkhuijsen72} map\footnote{We projected the map 
downloaded from {\tt http://www.mpifr-bonn.mpg.de/survey.html} into HEALPix pixelization.} at 0.820~GHz and the \citet{reich86} map at 1.4~GHz. 

At $10.9$, $12.7$, $14.7$ and $16.3$~GHz, similar frequencies to QUIJOTE, we use data from the COSMOSOMAS experiment \citep{watson05}. 
In order to minimize the $1/f$ noise, the data from this experiment was filtered by the suppression of the first seven harmonics in 
the FFT of the circular scans, which results in a flux loss on large angular scales. For this reason, the comparison with other 
experiments that preserve all the angular scales is not straightforward. It is necessary to account for the flux lost, as it was 
done in \citet{per20}. The fluxes presented in that paper are already corrected, so we directly take those fluxes.

We also use data from the 9-year release of the \wmap satellite\footnote{Downloaded from the LAMBDA database, {\tt http://lambda.gsfc.nasa.gov/}.} \citep{bennett13} 
to provide flux estimates at frequencies 23, 33, 41, 61 and 94~GHz. Recent data from the first release of the \planck mission\footnote{Downloaded from the 
\planck Legacy Archive, {\tt http://www.sciops.esa.int/index.php?project=planck\&page=}\\{\tt Planck\_Legacy\_Archive}.} \citep{cpp1} cover the 
frequencies 28, 44, 70, 100, 143, 217, 353, 545 and 857~GHz. We also download the released Type 1 CO maps \citep{cpp13}, which are used to correct the 
100, 217 and 353~GHz frequency maps from the contamination introduced by the CO rotational transition lines (1-0), (2-1) and (3-2), 
respectively. Finally, in the far-infrared spectral range we use Zodi-Subtracted Mission Average (ZSMA) COBE-DIRBE maps \citep{hauser98} 
at 240~$\mu$m (1249~GHz), 140~$\mu$m (2141~GHz) and 100~$\mu$m (2998~GHz).

\begin{figure*}
\centering
\includegraphics[width=12.5cm]{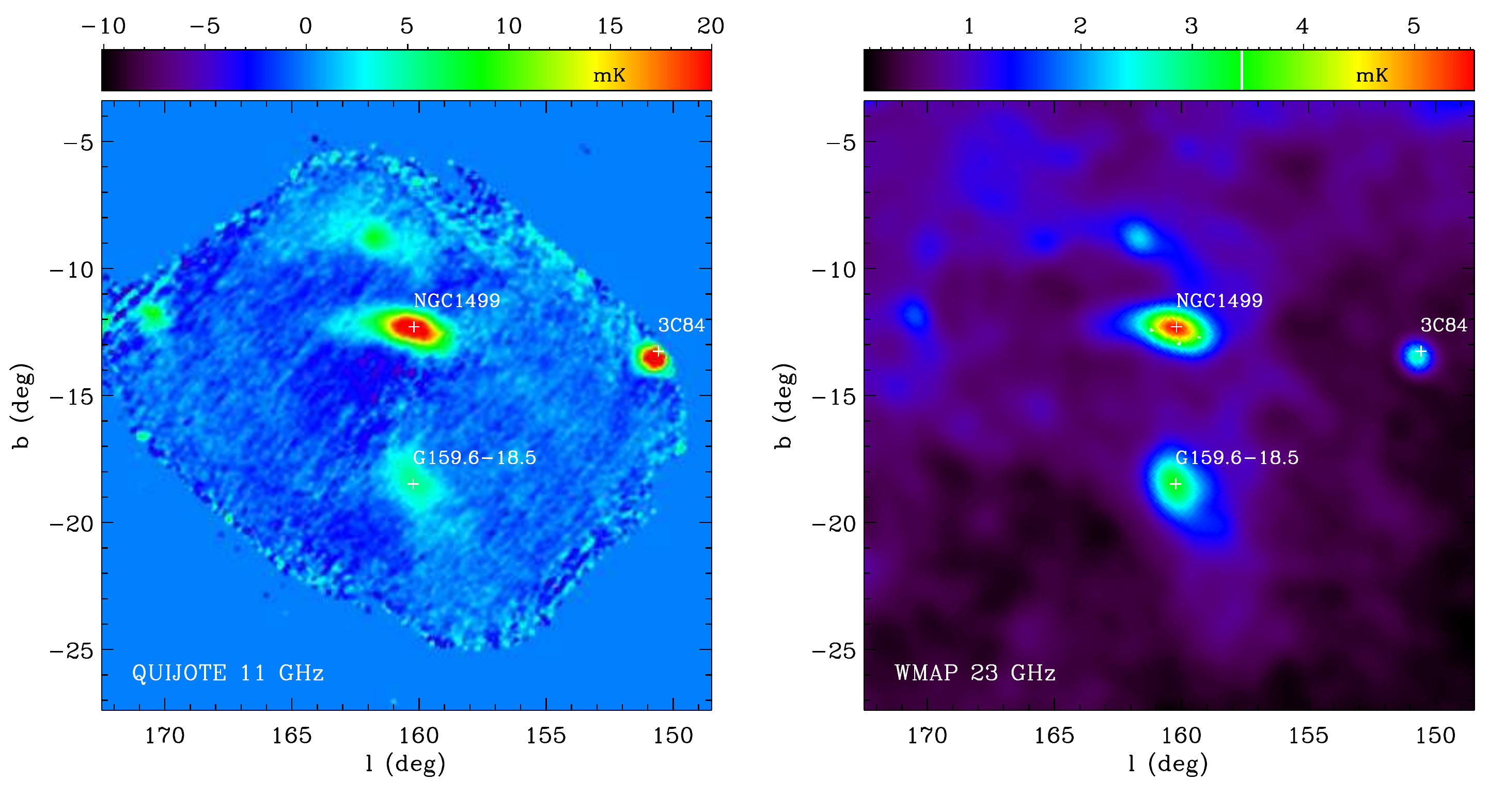}
\caption{\small QUIJOTE intensity map at 11 GHz of the whole region covered by the observations (left), in comparison with the \wmap 9-year 
map at 23~GHz (right). The positions of the G159.6-18.5 molecular cloud, the California HII region (NGC1499) and the 3C84 quasar are 
marked with crosses. The QUIJOTE map encompasses 277~deg$^2$, contains in total 149~h of observations, and its RMS is $\approx 80~\mu$K/beam. 
By comparing the relative amplitudes of California and G159.6-18.5 it can easily be noted how the presence of AME boosts the 
brightness of G159.6-18.5 at 23~GHz.}
\label{fig:map_bigperseus}
\end{figure*}

\begin{figure*}
\centering
\includegraphics[width=18.0cm]{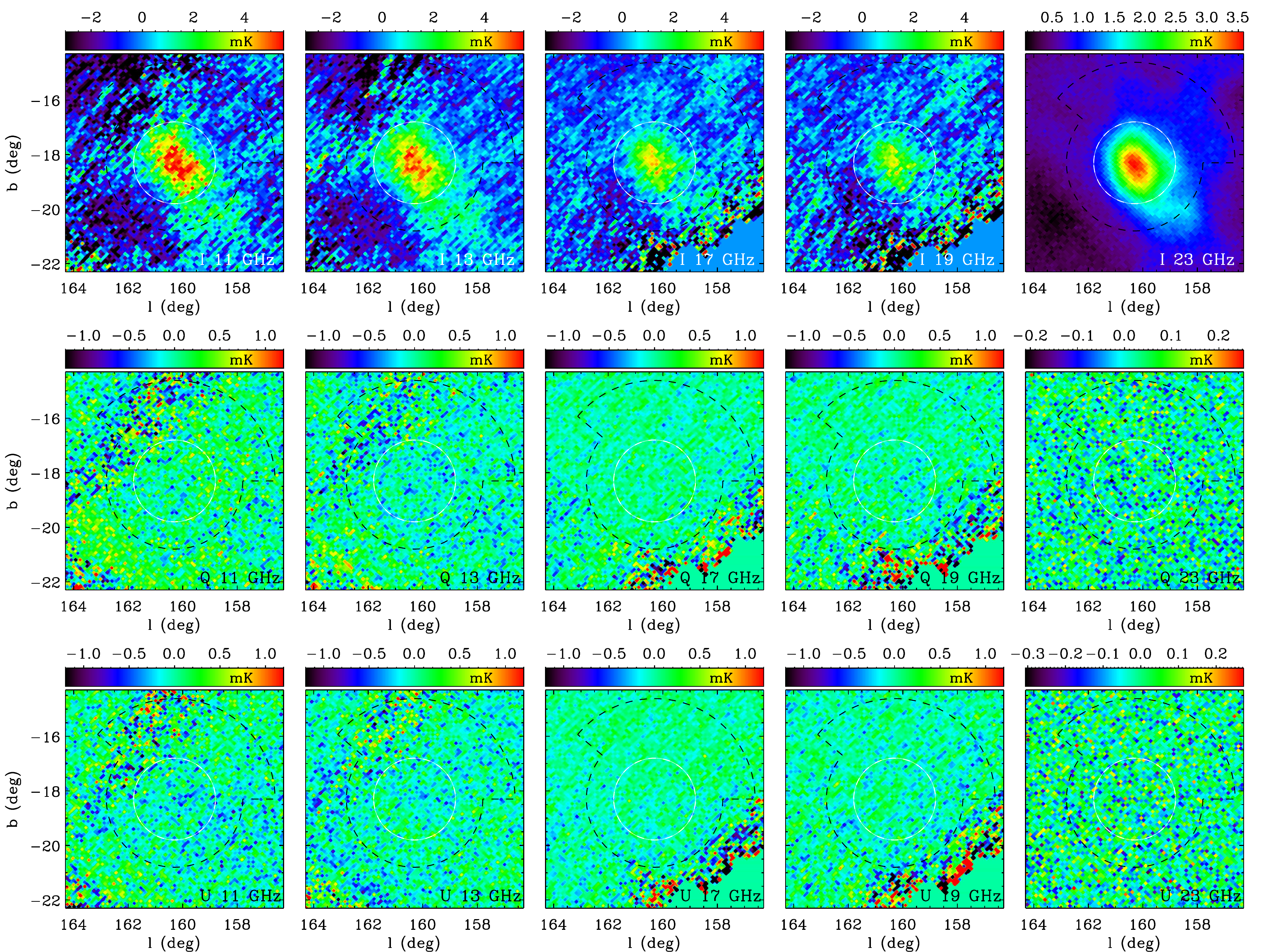}
\caption{\small Intensity and polarization maps at the QUIJOTE four frequency channels around the G159.6-18.5 molecular cloud. The Stokes 
Q and U maps show zero polarization. The solid circle shows the aperture we use for flux integration, whereas the dashed contour 
limits the extent of the ring we use for background subtraction. For comparison, in the last column we plot the corresponding \wmap 
9-year maps at 23~GHz.}
\label{fig:perseus_maps}
\end{figure*}

\begin{figure*}
\centering
\includegraphics[width=15.5cm]{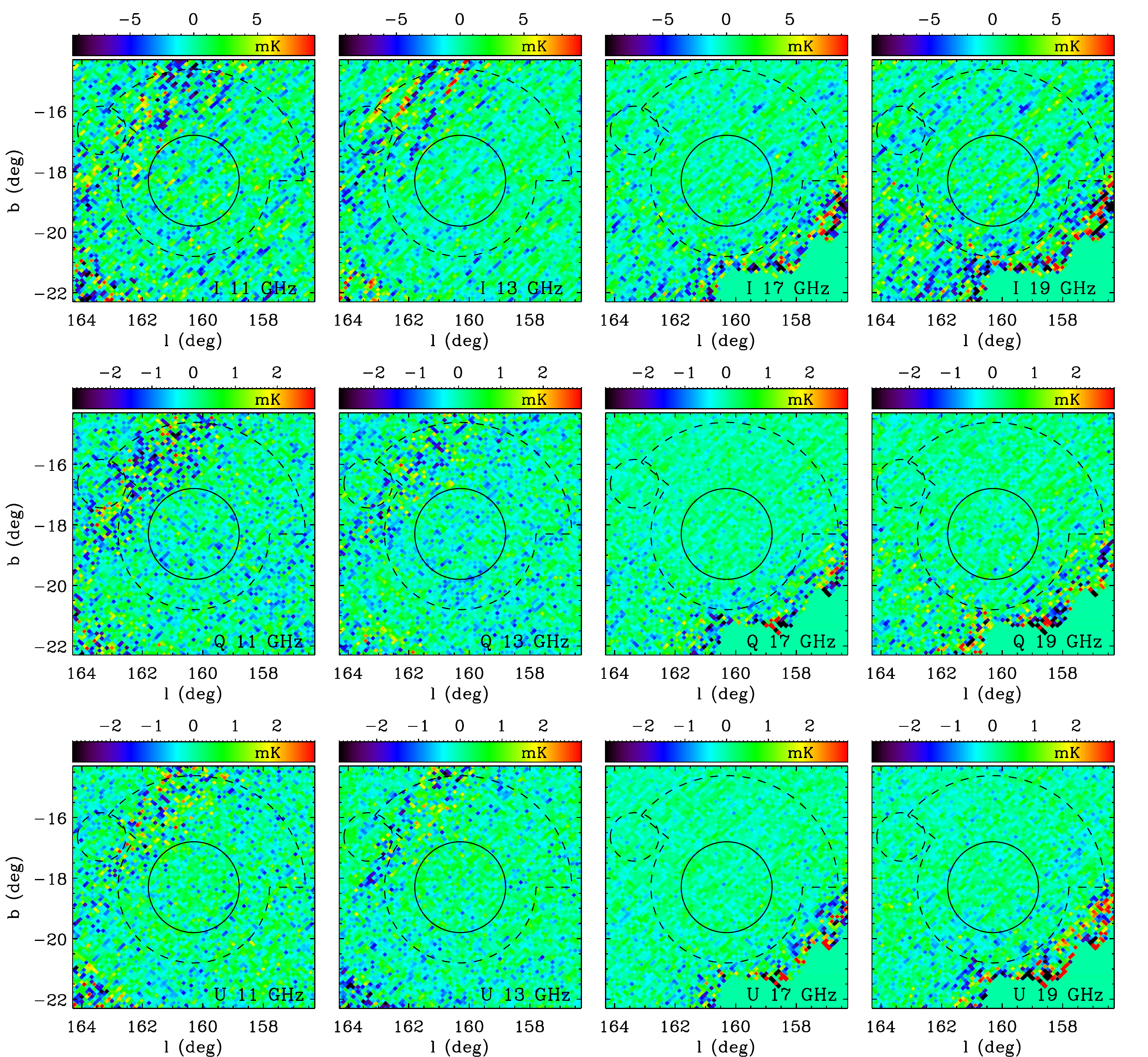}
\caption{\small Jackknife difference maps at the four frequencies around the G159.6-18.5 molecular cloud. The structure of this map is 
consistent with instrumental noise, which is predominantly Gaussian (the stripes are associated with regions with a higher noise 
due to a lower integration time per pixel - see text for details). In Table~\ref{tab:jk_tests} we show the 
pixel-to-pixel RMS values calculated in the background annulus enclosed by the two concentric circles of this plot and in the dashed circle.}
\label{fig:perseus_jk_maps}
\end{figure*}

\subsection{Methodology for flux estimation}\label{sec:methodology}

Intensity and polarization fluxes will be calculated by applying an aperture photometry integration on the maps.  This is a well known and widely used 
technique in this context  \citep{lopez11,dickinson11,genova11}, consisting of integrating temperatures of all pixels within a given aperture, and 
subtracting a background level calculated in an external ring. Instead of the mean, following \citet{per20}, we chose to use the median of all the 
pixels in the external ring as the estimate of the background level. The median is a better proxy for the real level in cases of strongly variable 
backgrounds with many outlier pixels. The flux is then given by
\begin{equation}
S_\nu = a(\nu)\left[\frac{\sum_{i=1}^{n1} T_i}{n_{\rm 1}} - {\tilde T}_j \right]~~,
\end{equation}
where $n_1$ is the number of pixels in the aperture, and $T_i$ and $T_j$ represent respectively the pixel thermodynamic temperatures 
in the aperture and in the background annulus. The median is calculated over the $n_2$ pixels in this annulus. The function $a(\nu)$ gives the
conversion factor from temperature to flux,
\begin{equation}
a(\nu) = \frac{h^2\nu^4}{2 k_{\rm b} T_{\rm cmb}^2 c^2}~{\rm sinh}^{-2}\left(\frac{h\nu}{2 k_{\rm b}T_{\rm cmb}}\right)n_1\Omega_{\rm pix}~~,
\label{eq:flux}
\end{equation}
where $h$ and $k_{\rm b}$ are the Planck and Boltzmann constants, $c$ the speed of light, $T_{\rm cmb}=2.725$~K \citep{mather99} the CMB 
temperature and $\Omega_{\rm pix}$ the solid angle subtended by each pixel.

The determination of the error associated with the previous estimate is crucial for the results of this paper. In a hypothetical case of perfect white 
uncorrelated noise, it could easily be estimated through:
\begin{equation}
\sigma(S_\nu) = a(\nu)\left[\frac{\sum_{i=1}^{n1}\sigma_i^2}{n_1^2}  + \frac{\pi}{2}\frac{\sum_{i=1}^{n2}\sigma_j^2}{n_2^2}\right]^{1/2}~~,
\label{eq:err1_flux}
\end{equation}
where $\sigma$ represents the error of each pixel.

However, in QUIJOTE the instrument noise is correlated due to the presence of $1/f$ residuals, and also the galactic background fluctuations introduce, 
mainly in intensity, an important contribution to the error which is correlated on the order of the beam size. Ideally we should 
then use the covariance matrices of the instrument and background noises. The former can be extracted through a characterization 
of the $1/f$ noise spectrum, however the latter is difficult to determine. Instead, in the previous equation we can  introduce in the 
denominator the number of independent pixels in the aperture and in the ring, which we will denote respectively as $n_1'$ and $n_2'$. 
The pixel variance will be calculated from the pixel-to-pixel standard deviation of all the pixel temperatures $T_j$ in the background, 
$\sigma(T_j)$. Obviously, the standard deviation of the pixels in the aperture would be biased by the presence of the source. On the contrary, 
the standard deviation of the pixels in the ring gives a reasonable estimate of the contributions of the background and of the instrumental 
noise to the true error. Therefore, the final equation that we will use to estimate errors in this paper reads as:
\begin{equation}
\sigma(S_\nu) = a(\nu)\sigma(T_j)\sqrt{\frac{1}{n_1'}+\frac{\pi}{2}\frac{1}{n_2'}}~~.
\label{eq:err2_flux}
\end{equation}

In the case of the error being completely dominated by the background, then we could use for $n_1'$ and $n_2'$ the number of beams 
in the aperture and in the background, $n_1^{\rm b}$ and $n_2^{\rm b}$. However, while being particularly strong in intensity, the background 
fluctuations from the Galactic emission are not so important in polarization. For this reason, in this case the relative contribution from the 
$1/f$ residuals to the uncertainty is significant. To quantify this, we selected 20 random positions around our source, G159.6-18.5, 
and performed flux integration on those positions using the same aperture and ring sizes. The standard deviation of those fluxes gives a reasonable estimation 
of the true noise of our flux estimate. From this analysis we determined that for intensity $n_{1,2}'=n_{1,2}^{\rm b}$, while for polarization 
$n_{1,2}'=2 n_{1,2}^{\rm b}$. This is what we will use in our estimation of the flux errors.

In cases of low signal-to-noise fluxes, or when placing upper limits on the polarized flux $P$, as it will be our case, it is necessary to de-bias the 
fluxes derived from the aperture photometry integration. This requirement comes from the fact that the posterior distribution of the polarized 
intensity $P$ does not follow a normal (Gaussian) distribution. Furthermore $P$ is a quantity that must always be positive, and this introduces a 
bias into any estimate. For any true $P_0$ we would expect to measure on average a polarization $P>P_0$. In order to get the de-biased fluxes, 
$P_{\rm db}$, from the measured ones, $P$, we choose the  Bayesian approach described in \citet{vaillancourt06} and in \citet{rubino12b}, consisting of integrating the 
analytical posterior probability density function over the parameter space of the true polarization. The same posterior can not be used for the 
polarization fraction, $\Pi=P/I\times 100$, as it follows a different distribution. As, to our knowledge, there is not in the literature any analytical 
solution for the posterior distribution of $\Pi$, we numerically evaluate this function by applying Monte-Carlo simulations. This approach has 
already been carried out in \citet{lopez11} and in \citet{dickinson11}.

\begin{figure*}
\centering
\includegraphics[width=13.5cm]{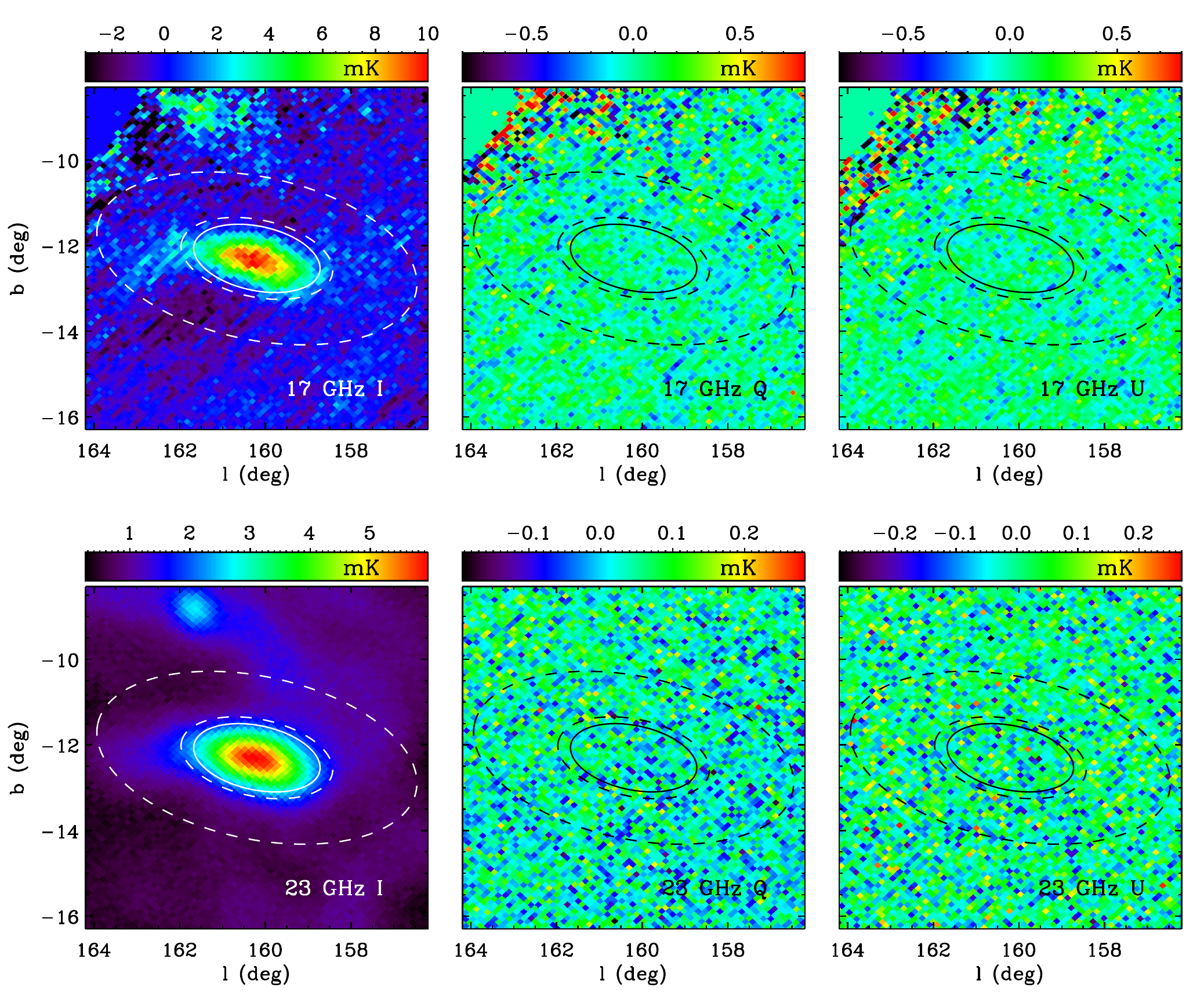}
\caption{\small Intensity and polarization maps around the California HII region, at 17 GHz from QUIJOTE and at 23 GHz from \wmap 
9-year data. The Stokes Q and U maps show zero polarization as is expected for a free-free dominated region. This test therefore 
demonstrates the absence of significant polarization systematics inherent to our experiment or to the data processing. The 
ellipses show the aperture and the background annulus that we use to derive polarization upper limits in this region.}
\label{fig:california_maps}
\end{figure*}

\begin{table*}
\begin{center}
\begin{tabular}{ccccccccccccccc}
\hline\hline
\noalign{\smallskip}
$\nu$ (GHz) &~& \multicolumn{3}{c}{$\sigma_{\rm I}$ ($\mu$K/pixel)}&~&  \multicolumn{3}{c}{$\sigma_{\rm Q}$ ($\mu$K/pixel)}&~&   \multicolumn{3}{c}{$\sigma_{\rm U}$ ($\mu$K/pixel)}  &~& $\sigma_{\rm Q,U}$ ($\mu$K~s$^{1/2}$)  \\
\noalign{\smallskip} 
\cline{3-5}\cline{7-9}\cline{11-13}\cline{15-15}
\noalign{\smallskip} 
 & ~& Map   & Sum  &  Diff. &~& Map  &  Sum  &  Diff. &~& Map & Sum  &  Diff. &~& Map\\
\noalign{\smallskip}\hline\noalign{\smallskip}
$11.2$  &&   $1592$  &  $1593$ &  $1277$  &&	  $389$   &   $396$  & $394$  &&       $361$  &  $361$  & $361$  && 1478\\
$12.9$  &&   $1328$  &  $1377$ &  $1119$  &&	  $300$   &   $308$  & $314$  &&       $293$  &  $301$	& $297$  && 1179\\
$16.7$  &&   $ 861$  &  $ 861$ &  $ 755$  &&	  $167$   &   $169$  & $171$  &&       $166$  &  $166$	& $161$  && 1158\\
$18.7$  &&   $1040$  &  $1041$ &  $ 966$  &&	  $199$   &   $202$  & $213$  &&       $201$  &  $205$	& $201$  && 1461\\
\noalign{\smallskip}\hline\noalign{\smallskip}
$11.2$  &&   $1377$  &  $1415$ &  $1320$  &&	  $406$   &   $418$  & $460$  &&       $315$  &  $335$  & $346$  && 1283\\
$12.9$  &&   $1331$  &  $1512$ &  $1529$  &&	  $352$   &   $387$  & $349$  &&       $309$  &  $339$  & $351$  && 1192\\
$16.7$  &&   $ 672$  &  $ 673$ &  $ 651$  &&	  $142$   &   $142$  & $140$  &&       $125$  &  $125$  & $136$  && 1009\\
$18.7$  &&   $ 803$  &  $ 809$ &  $ 876$  &&	  $149$   &   $150$  & $157$  &&       $164$  &  $165$  & $163$  && 1413\\
\noalign{\smallskip}
\hline\hline
\end{tabular}
\end{center}
\normalsize
\medskip
\caption[tab:jk_tests]{RMS pixel-to-pixel (pixel size $6.9$~arcmin) calculated in the original maps, and in the average and difference 
divided by two of the two jackknife maps, in two different regions. The values above and below the horizontal line have been calculated 
respectively in the background annulus and in the dashed circle depicted in Fig.~\ref{fig:perseus_jk_maps}. The numbers in the last 
column correspond to the average noises in the original $Q$ and $U$ maps normalized by the integration time per pixel (units: $\mu$K~s$^{1/2}$).}
\label{tab:jk_tests}
\end{table*}

\section{Results and discussion}

\subsection{Maps and consistency tests}\label{sec:maps}

In Fig.~\ref{fig:map_bigperseus} we show the intensity map at 11~GHz resulting from combining 149~h of observations, where emissions 
from G159.6-18.5, the California nebula (NGC1499) and the 3C84 quasar are clearly visible. More detailed $I$, $Q$ and $U$ maps at our four 
frequencies around the position of G159.6-18.5 are shown in Fig.~\ref{fig:perseus_maps}. The $Q$ and $U$ maps on this source are 
consistent with zero polarization, and therefore upper limits on the polarized intensity will be extracted in section~\ref{sec:pol_limits}. 
Some stripping is clearly visible in the maps, which is produced by the presence of regions with a higher noise due to 
a lower integration time per pixel and, to a lesser extent, to $1/f$ residuals. The inhomogenities in the coverage (integration time per pixel) 
maps are caused by the separation of the horns in the focal plane, which leads the sky coverage to be different when we observe the 
field before or after crossing the local meridian. In the $Q$ and $U$ maps at 11 and 13~GHz of Fig.~\ref{fig:perseus_maps}, the two 
orthogonal stripes with clearly higher noise correspond to regions with integration of $\sim 3-7$~s/pixel (pixel size 6.9~arcmin). 
By comparison, in the central region inside the circle where we perform the aperture photometry, the integration time is $\sim 30-35$~s/pixel, 
resulting in a lower pixel-to-pixel dispersion of the data.

An important consistency test, that may reveal the presence of systematics and other spurious effects, is obtained through jackknife maps. 
We have uniformly split our full dataset in two halves in such a way that the maps of number of hits associated to these two halves are as 
similar as possible. The differences of the two halves divided by two, for the intensity and polarization maps at our four frequencies, are 
shown in Fig.~\ref{fig:perseus_jk_maps}. As expected, the intensity emission coming from G159.6-18.5 is consistently cancelled out 
in these maps. While a similar striping pattern to the maps in Fig.~\ref{fig:perseus_maps} is seen, these maps are dominated by instrumental 
noise. This is confirmed by the noise values shown in Table~\ref{tab:jk_tests}, where we compare the pixel-to-pixel RMS calculated in two 
different regions: the external ring that we will use for background subtraction when calculating the intensity and polarization fluxes, and 
a region of very low sky emission enclosed by the dashed circle represented in Fig.~\ref{fig:perseus_jk_maps}. The RMS values 
in $Q$ and $U$ are similar in the original maps and in the sum and difference of the two halves. They are typically $\sim 250~\mu$K/pixel 
(pixel size: 6.9~arcmin) or $\sim 25~\mu$K/beam (beam size: $1\sim $~degree). In the case of the $I$ maps, the 
RMS in the background ring are higher in the sum than in the difference because of the emission of the source. In the circle with low sky 
signal the values in the sum and in the difference maps are very similar.

 The last column of Table~\ref{tab:jk_tests} show the average RMS levels in the $Q$ and $U$ maps normalized by the integration time per pixel. 
As the number of hits per pixel is very inhomogeneous, to calculate these numbers we have made a realization of Gaussian noise in which we 
assign to each pixel a noise proportional to $t_{\rm pix}^{-1/2}$, where $t_{\rm pix}$ is the integration time per pixel, and then calculate 
the pixel-to-pixel RMS. The amplitudes of the white-noise in the spectra of the time-ordered-data range between 898~$\mu$K~s$^{1/2}$ 
(at 16.7~GHz) and 1228~$\mu$K~s$^{1/2}$ (at 11.2~GHz)\footnote{Note that in this article we are using data from only two out of the four 
horns of QUIJOTE. Then, the global sensitivities of the experiment are a factor $\sqrt{2}$ better, i.e. $\approx 650~\mu$K~s$^{1/2}$, the 
number quoted in Section~\ref{sec:quijote_data}}. $1/f$ residuals make the noises calculated on the maps only slightly higher, 
typically by a factor $\sim 15\%$, confirming our previous statement that these maps are dominated by white (Gaussian) noise.

Another important consistency test for the presence of systematics, and in particular for the $I$ to $Q/U$ polarization leakage, is to confirm 
that our polarization maps are consistent with noise in the position of unpolarized sources. This verification is provided by the nearby 
California HII region, which is also covered by our observations. As any standard HII region, it is dominated by free-free emission at the 
QUIJOTE frequencies, which is known to be practically unpolarized. In Fig.~\ref{fig:california_maps} we show intensity and polarization 
maps towards the California region, showing that the $Q$ and $U$ maps are consistent with zero polarization.

\subsection{Intensity SED}\label{sec:intensity_sed}

\begin{table} 
\begin{center}
\begin{tabular}{cccc}
\hline\hline
\noalign{\smallskip}
Frequency & Flux & Flux density & Telescope/\\
(GHz)     & density (Jy)        &  residual (Jy) & survey \\
\noalign{\smallskip}\hline\noalign{\smallskip} 
$  0.408$  &   $  10.5\pm  4.0$    &	 $  2.7\pm   4.2$  & Haslam \\
$   0.82$  &   $   7.4\pm  2.1$    &	 $  0.1\pm   2.4$  & Dwingeloo  \\
$   1.42$  &   $   6.8\pm  1.6$    &	 $ -0.1\pm   2.0$  & Reich  \\
$  10.9 $  &   $  16.1\pm  1.8$    &	 $ 10.3\pm   1.9$  & COSMO.  \\
$  11.2 $  &   $  15.0\pm  2.0$    &	 $  9.2\pm   2.1$  & QUIJOTE  \\
$  12.7 $  &   $  20.0\pm  2.2$    &	 $ 14.2\pm   2.2$  & COSMO.  \\
$  12.9 $  &   $  18.1\pm  2.1$    &	 $ 12.4\pm   2.1$  & QUIJOTE  \\
$  14.7 $  &   $  28.4\pm  3.1$    &	 $ 22.6\pm   3.1$  & COSMO. \\
$  16.3 $  &   $  35.8\pm  4.0$    &	 $ 30.0\pm   4.0$  & COSMO.  \\
$  16.7 $  &   $  33.9\pm  2.4$    &	 $ 28.1\pm   2.5$  & QUIJOTE  \\
$  18.7 $  &   $  35.2\pm  3.7$    &	 $ 29.2\pm   3.8$  & QUIJOTE  \\
$  22.8 $  &   $  40.2\pm  2.4$    &	 $ 34.0\pm   2.5$  & \wmap  \\
$  28.4 $  &   $  40.4\pm  2.4$    &	 $ 33.5\pm   2.6$  & \planck  \\
$  33.0 $  &   $  38.1\pm  2.4$    &	 $ 30.4\pm   2.8$  & \wmap  \\
$  40.7 $  &   $  32.8\pm  2.5$    &	 $ 23.1\pm   3.3$  & \wmap  \\
$  44.1 $  &   $  29.8\pm  2.6$    &	 $ 19.1\pm   3.7$  & \planck  \\
$  60.8 $  &   $  27.5\pm  3.8$    &	 $  8.3\pm   6.9$  & \wmap  \\
$  70.4 $  &   $  32.3\pm  4.9$    &	 $  5.2\pm  10.0$  & \planck  \\
$  93.5 $  &   $  59.5\pm  9.3$    &	 $  1.8\pm  22.3$  & \wmap  \\
$ 100	$  &   $    81\pm   17$    &	 $   11\pm    10$  & \planck  \\
$ 143	$  &   $   194\pm   24$    &	 $  -17\pm    82$  & \planck  \\
$ 217	$  &   $  1011\pm  122$    &	 $  196\pm   320$  & \planck  \\
$ 353	$  &   $  4286\pm  446$    &	 $  344\pm  1376$  & \planck \\
$ 545	$  &   $ 14858\pm 1470$    &	 $  208\pm  4588$  & \planck \\
$ 857	$  &   $ 45235\pm 4045$    &	 $-1352\pm 13168$  & \planck \\
$1249	$  &   $ 86696\pm 6674$    &	 $-4878\pm 25315$  & DIRBE  \\
$2141	$  &   $114650\pm 6891$    &	 $ 6845\pm 43590$  & DIRBE  \\
$2998	$  &   $ 54361\pm 2624$    &	 $ -837\pm 28264$  & DIRBE  \\
\noalign{\smallskip}
\hline\hline
\end{tabular}
\end{center}
\normalsize
\medskip
\caption[tab:fluxes_sed]{Flux densities for G159.6-18.5 in the Perseus molecular cloud. All fluxes have been calculated through aperture 
photometry in a ring of radius $1.7^\circ$ and subtracting the median of the background in a ring between $1.7^\circ$ and $1.7\sqrt{2}^\circ$, 
except those coming from the COSMOSOMAS experiment (10.9, 12.7, 14.7 and 16.3~GHz), which have been taken from \citet{per20}. Also shown 
are the residual AME fluxes, obtained after subtraction of the free-free, CMB and thermal dust components. The last column indicate 
the telescope or survey from which the data have been extracted.}
\label{tab:fluxes_sed}
\end{table}

As was indicated in section~\ref{sec:ancillary}, we take the COSMOSOMAS fluxes for G159.6-18.5 from \citet{per20}, which have 
already been corrected for the flux loss caused by the filtering of COSMOSOMAS data. While here we will use aperture 
photometry to derive our fluxes, in \citet{pip25} they were obtained by fitting the amplitude of an elliptical Gaussian with a 
fixed size of $1.6^\circ\times 1.0^\circ$ (FWHM). In a first-order approximation, the fluxes obtained through Gaussian fitting 
will be equivalent to those obtained from aperture photometry using a given aperture size. Therefore, in order to get a reliable 
intensity SED we choose a size for the aperture that gives the most similar fluxes to those presented in \citet{per20} for the 
\citet{haslam82}, \citet{berkhuijsen72}, \cite{reich86}, \wmap, \planck and DIRBE maps (it must be noted that the \wmap and \planck maps used 
in this work correspond to a different release to that used in \citet{per20}, but this will have a negligible effect). After trying 
different values, we found that a radius of $1.7^\circ$ gives the best agreement, with a very low reduced chi-squared of 
$\chi^2_{\rm red}=0.098$. The median of the background is computed in an external ring between $1.7^\circ$ and $1.7\sqrt{2}^\circ$, 
which has the same area as the aperture. The derived fluxes in the QUIJOTE maps and in the other ancillary maps are listed in 
Table~\ref{tab:fluxes_sed}.

\begin{figure}
\centering
\includegraphics[width=\columnwidth]{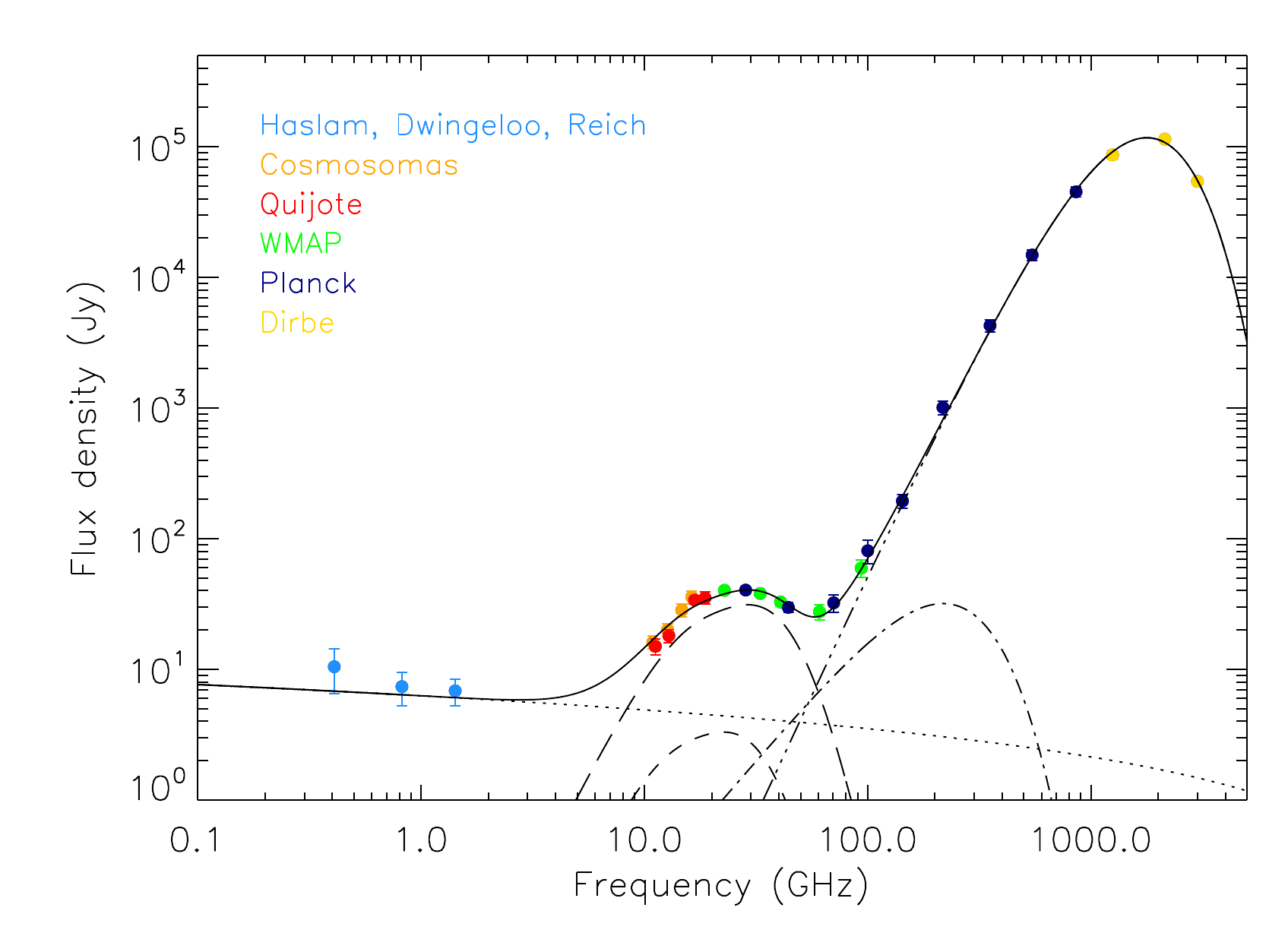}
\caption{\small Spectral energy distribution of G159.6-18.5. QUIJOTE data points are depicted in red, together with 
other ancillary data including COSMOSOMAS, \wmap 9-year data and \planck data. At intermediate frequencies the excess emission 
associated with the AME clearly shows up. A joint fit has been performed to all the data points except 100 and 217~GHz (which are 
affected by CO contamination), consisting in the following components: free-free (dotted line), spinning dust from a mixed 
environment with a high-density molecular (long-dashed line) and a low-density atomic (dashed line) gas, CMB (dash-dotted line) 
and thermal dust (dashed-triple-dotted line). The solid line represents the sum of all the components.}
\label{fig:sed}
\end{figure}

The final SED is depicted in Fig.~\ref{fig:sed}, where the presence of AME clearly shows up at intermediate frequencies as an excess 
of emission over the other components. The intensities derived from these QUIJOTE observations trace, for the first time after the original 
measurements of the COSMOSOMAS experiment \citep{watson05}, the downturn of the spectrum at frequencies below $\sim 20$~GHz, 
as it is predicted by spinning dust models. In total, 13 data points are dominated by AME: the four QUIJOTE points, the four 
COSMOSOMAS points, the \wmap $22.8$, $33.0$ and $40.7$~GHz frequencies and the \planck $28.4$ and $44.1$~GHz frequencies. 
We perform a joint multi-component fit to all the data, consisting of free-free emission, 
which dominates in the low-frequency tail, spinning dust, which dominates the intermediate frequencies, a CMB component, and thermal dust, 
which dominates the high-frequency end. As it was done in \citet{per20}, to avoid possible CO residuals we exclude from the fit 
the 100~GHz and the 217~GHz values. We fix the spectrum of the free-free using the standard formulae shown in \citet{per20}, with a 
value for the electron temperature typical of the solar neighbourhood, $T_{\rm e}=8000$~K, and fit for its amplitude, which is 
parameterized through the emission measure ($EM$). 

Following \citet{per20}, we consider a high-density molecular 
phase and a low-density atomic phase which produce spinning dust emission, and fit their respective amplitudes, which are given by the 
hydrogen column densities $N_{\rm H}^{\rm mol}$ and $N_{\rm H}^{\rm at}$. The CMB amplitude is denoted by $\Delta T_{\rm cmb}$, and would 
correspond to the average of the primordial CMB fluctuations in the aperture. Finally, 
the thermal dust is modelled as a single-component modified blackbody curve, 
$\tau_{\rm 250}(\nu/1200~{\rm GHz})^{\beta_{\rm d}} B_{\nu}(T_{\rm d})$, which depends on three parameters: the optical depth at $250~\mu$m 
($\tau_{\rm 250}$), the emissivity spectral index ($\beta_{\rm d}$) and the dust temperature ($T_{\rm d}$). Therefore, we jointly fit 
7 parameters to all the data points: $EM$, $N_{\rm H}^{\rm mol}$, $N_{\rm H}^{\rm at}$, $\Delta T_{\rm cmb}$, $\tau_{\rm 250}$, 
$\beta_{\rm d}$ and $T_{\rm d}$.

To define the spinning dust spectra of the molecular and atomic phases we use the {\sc spdust.2} 
code\footnote{http://www.tapir.caltech.edu/$\sim$yacine/spdust/spdust.html} \citep{ali09,silsbee11}. Initially we use the same values as in 
\citet{per20} for the different parameters on which the spinning dust emissivity depends. This involves a modification of the code, as by 
default it uses $a_{0,1}=0.35$~nm and $a_{0,2}=3.0$~nm for the centroids of the two lognormal functions defining the dust grains size 
distributions, while the carbon abundance per hydrogen nucleus, $b_{\rm C}$, is selected from any of the values in table~1 of 
\citet{weingartner01}. Following \citet{per20}, we use instead $a_{0,1}=0.58$~nm and $a_{0,1}=0.53$~nm, respectively for the molecular 
and atomic phases, and $b_{\rm C}=68$~ppm. Using these two models, we get a good fit to our full dataset, with $\chi^2_{\rm red}=0.99$, 
where the spinning dust component is clearly dominated by the molecular phase, as was found by \citet{per20}.

To analyze the possibility of the existence of slightly different spinning dust models that could provide a better fit to the data, we take the 
AME residual fluxes from the previous fit, and produce a grid of models varying some of the parameters of the molecular phase component. 
In particular, we vary: i) the hydrogen number density $n_{\rm H}$ between 10 and 500~cm$^{-3}$ with a step of 5~cm$^{-3}$; ii) the kinetic 
gas temperature $T_{\rm g}$ between 5 and 200~K in steps of 5~K; iii) the intensity of the radiation field relative to the average interstellar 
radiation field, for which we consider only the values $G_0=1$ and 2; iv) and the hydrogen ionization fraction, for which we consider the 
values $x_{\rm H}=10$, $112$, $1000$ and $10000$~ppm. The best fit is obtained for $G_0=1$ and 
$x_{\rm H}=112$~ppm, the same values used in \citet{per20}. As the fit is very degenerate, to define the most-likely values for 
$n_{\rm H}$ and $T_{\rm g}$ we set Gaussian priors on four different parameters. First, we put soft priors on $n_{\rm H}$ and 
$T_{\rm g}$ centred on the same values used in \citet{per20}, $(n_{\rm H})_0=250$~cm$^{-3}$ and $(T_{\rm g})_0=40$~K, and with 
standard deviations $\sigma(n_{\rm H})=80$~cm$^{-3}$ and $\sigma(T_{\rm g})=60$~K. An additional prior on $N_{\rm H}^{\rm mol}$ can 
be derived from the canonical relation $2.13\times 10^{24}$~H~cm$^{-2}=1\tau_{100}$ \citep{finkbeiner04}. Using $\tau_{250}$ and 
$\beta_{\rm d}$ from our best-fit model of the thermal dust component, we extrapolate the optical depth to $100~\mu$m, and find 
$N_{\rm H}^{\rm mol}=2.909\times 10^{21}$~H~cm$^{-2}$. We use this value to define the centre of the Gaussian prior, and 
$\sigma(N_{\rm H}^{\rm mol})=2\times 10^{21}$~H~cm$^{-2}$. Finally, the ratio between the hydrogen column density and the hydrogen 
volume density, $z=N_{\rm H}^{\rm mol}/n_{\rm H}$, gives an estimate of the length along the line of sight of the 
spinning-dust-emitting region. We assume that this length might be of the order of the transverse angular size of the source. The 
source subtends an angle of around $2^\circ$, which at the distance of the Perseus complex, 260~pc \citep{cernicharo85}, corresponds 
to 9.08~pc. We therefore set a fourth prior on this quantity defined by $z_0=9.08$~pc and $\sigma(z)=4$~pc.

\begin{figure*}
\centering
\includegraphics[width=13cm]{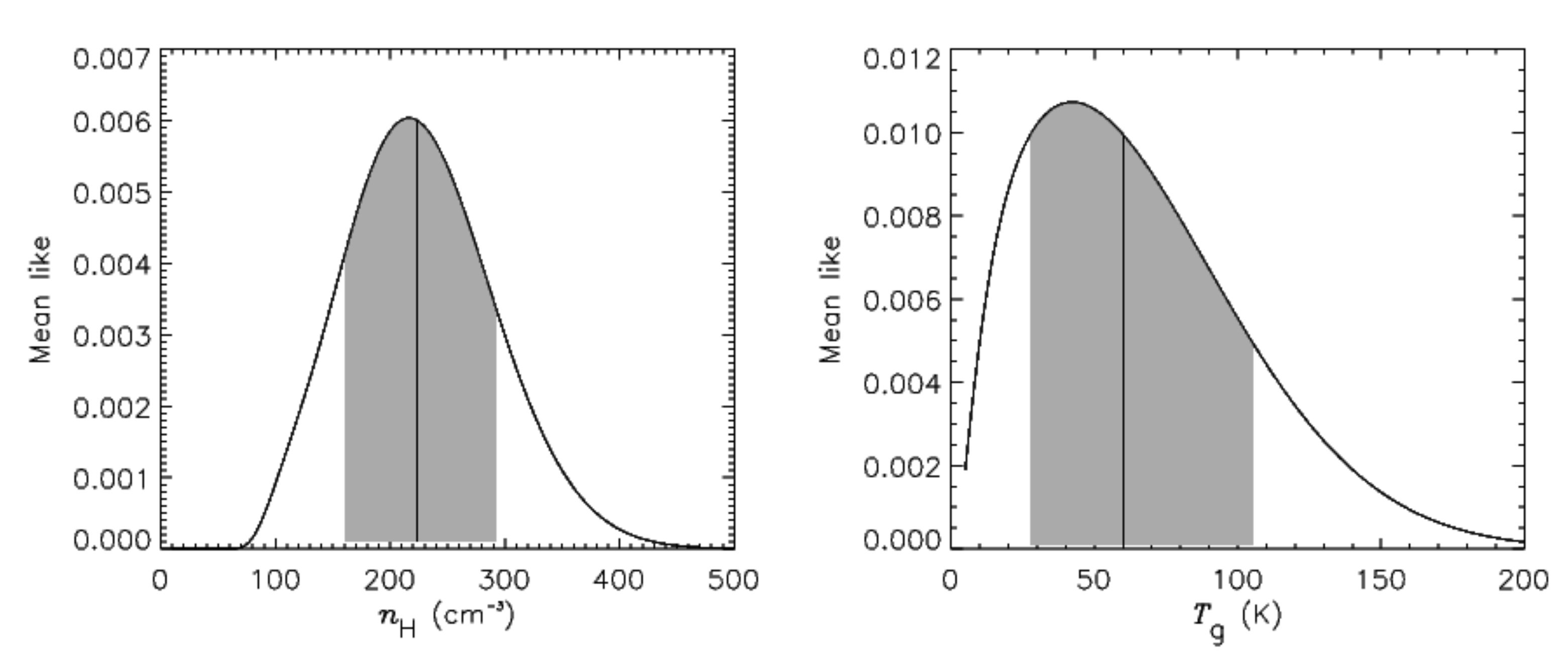}
\caption{\small Likelihood distributions for the hydrogen number density (left) and for the kinetic gas temperature (right) associated with the 
molecular phase of the spinning dust component. These were obtained after marginalizing over the rest of parameters on which the spinning dust 
emissivity depends (see text for details). The vertical lines show the most likely value, defined through the 50\% integral of the cumulative 
probability distribution, and the dashed regions depict the 68\% confidence interval around this value.}
\label{fig:marglike}
\end{figure*}

In Fig.~\ref{fig:marglike} we show the marginalized likelihoods over $n_{\rm H}$ and $T_{\rm g}$. We define the best-values for these 
parameters from the 50\% integrals of the probability distribution, and the confidence intervals from the region encompassing
the 68\% of the area around those values. We get $n_{\rm H}=223.2^{+69.5}_{-62.8}$ and $T_{\rm g}=60.2^{+45.3}_{-32.6}$~K. As it was said before, the values of 
the intensity of the radiation field and of the hydrogen ionization fraction that maximize the likelihood are $G_0=1$ and 
$x_{\rm H}=112$~ppm, respectively. We fix the other parameters of the molecular phase, and all the parameters corresponding 
to the atomic phase, to the same values that were used in \citet{per20}. All these values are shown in Table~\ref{tab:sed_fit_params}. 
In this table, $x_{\rm C}$ represents the ionized carbon fractional abundance, $y$ the molecular hydrogen fractional abundance, 
and $\beta$ the average dipole moment per atom. The meaning of the other parameters have been explained before in the text.
We then obtain the corresponding spinning dust spectra for 
the molecular and atomic phases using these parameters as inputs for {\sc spdust.2}. Fixing these spectra, we perform a joint fit 
of the five aforementioned components, obtaining the best-fit values for the 7 parameters defining these models, which are also 
shown in Table~\ref{tab:sed_fit_params}. We get $\chi^2_{\rm red}=0.99$, the same value as before, so we do not manage to improve 
the quality of the global fit after improving the spinning dust models. This highlights the difficulty of constraining the parameters 
on which the spinning dust emission depends due to the the strong degeneracies between them.

\begin{table} 
\begin{center}
\begin{tabular}{lcc}
\hline\hline
\noalign{\smallskip}
$EM$ (cm$^{-6}$pc) & \multicolumn{2}{c}{$23.9\pm 1.9$} \\
\hline
\noalign{\smallskip}
& Molecular & Atomic \\
\hline
\noalign{\smallskip}
$n_{\rm H}$ (cm$^{-3}$)  & $223.2$  & 30  \\
$G_0$  &   1  & 2 \\
$T_{\rm g}$ (K)      & $60.2$   & 100 \\
$x_{\rm H}$ (ppm)      & 112  & 410  \\
$x_{\rm C}$ (ppm)      & 1    & 100  \\
$y$                    & 1    & 0.1  \\
$a_{0,1}$ (nm)        & $0.58$  &  $0.53$  \\
$a_{0,2}$ (nm)        & $3.0$  &  $3.0$  \\
$b_{\rm C}$ (ppm)     &  68    &  68  \\ 
$\beta$ (D)          & 9.34 & 9.34 \\
\noalign{\smallskip}
$N_{\rm H}$ (10$^{21}$ cm$^{-2}$) & $3.88\pm 0.52$ &   $0.52\pm 0.65$ \\
\noalign{\smallskip}
\hline
\noalign{\smallskip}
$\Delta T_{\rm cmb}$ ($\mu$K) & \multicolumn{2}{c}{$22.6\pm 13.6$}\\
\hline
\noalign{\smallskip}
$\beta_{\rm d}$ & \multicolumn{2}{c}{1.73 $\pm$ 0.11}  \\
$T_{\rm d}$ (K) & \multicolumn{2}{c}{18.2 $\pm$ 0.6}\\
$\tau_{250}$ & \multicolumn{2}{c}{($2.78\pm 0.41$)$\times$10$^{-4}$}  \\
\hline
\noalign{\smallskip}
$\chi^2_{\rm red}$ & \multicolumn{2}{c}{0.99}\\
\noalign{\smallskip}
\hline\hline
\end{tabular}
\end{center}
\caption[tab:sed_fit_params]{Model parameters. The emission measure $EM$ is fitted to the data and defines the amplitude of the 
free-free emission. Parameters from $n_{\rm H}$ to $\beta$ (see section~\ref{sec:intensity_sed} for an explanation of their meaning) 
are used as inputs for {\sc spdust.2}, and define the shape of the spinning dust spectra for the molecular and atomic phases. The 
amplitude of these spectra are determined through the fit to the data, and are given by the hydrogen column density $N_{\rm H}$. 
The best-fit values for the amplitude of the CMB component $\Delta T_{\rm cmb}$ and for the three parameters defining the thermal 
dust spectrum, as well as the reduced chi-squared, are also shown.}
\label{tab:sed_fit_params}
\end{table}

The total hydrogen column density is $(4.40\pm 0.83)\times 10^{21}$~H~cm$^{-2}$. This is a 
bit higher than the expected value of $2.89\times 10^{21}$~H~cm$^{-2}$, which has been derived from the aforementioned 
$\tau_{100}-N_{\rm H}$ canonical relation, and extrapolating $\tau_{100}$ to $\tau_{250}$ using the fitted spectrum for the 
thermal dust emission. The inferred lengths of the two spinning dust phases along the line of sight, $z=N_{\rm H}/n_{\rm H}$, are 
respectively $z^{\rm mol}=5.63\pm 0.76$~pc and $z^{\rm at}<12.6$~pc (a 68.3\% C.L. upper bound is used here, as the 
error bar is higher than the estimate). These values are of the order, or compatible, with the transverse 
size of the region, $9.08$~pc. Our fitted values for $\beta_{\rm d}$ and $T_{\rm d}$ are consistent within 1-sigma with those derived in \citet{per20}. 
On the other hand, we get lower values for $EM$, $N_{\rm H}$ and $\tau_{250}$, but this is because these depend on the solid angle 
subtended by the region.  This value is a factor $\approx 5$ higher in our case, because we are using aperture photometry instead of 
Gaussian fitting. When this factor is taken into account, then our values are brought into a better agreement with those of 
\citet{per20}. Finally, we find a positive value for $\Delta T_{\rm cmb}$, whereas in \citet{per20} it was negative. However, we have 
checked that our value agrees with the average level of the CMB anisotropies within the aperture, which has been found to be 
$23.6~\mu$K in the {\it Planck}-DR1 CMB map resulting from the SMICA component separation method\footnote{Downloaded from the 
\planck Legacy Archive, {\tt http://www.sciops.esa.int/index.php?project=planck\&page=}\\{\tt Planck\_Legacy\_Archive}.} \citep{cpp12}.

\begin{figure}
\centering
\includegraphics[width=\columnwidth]{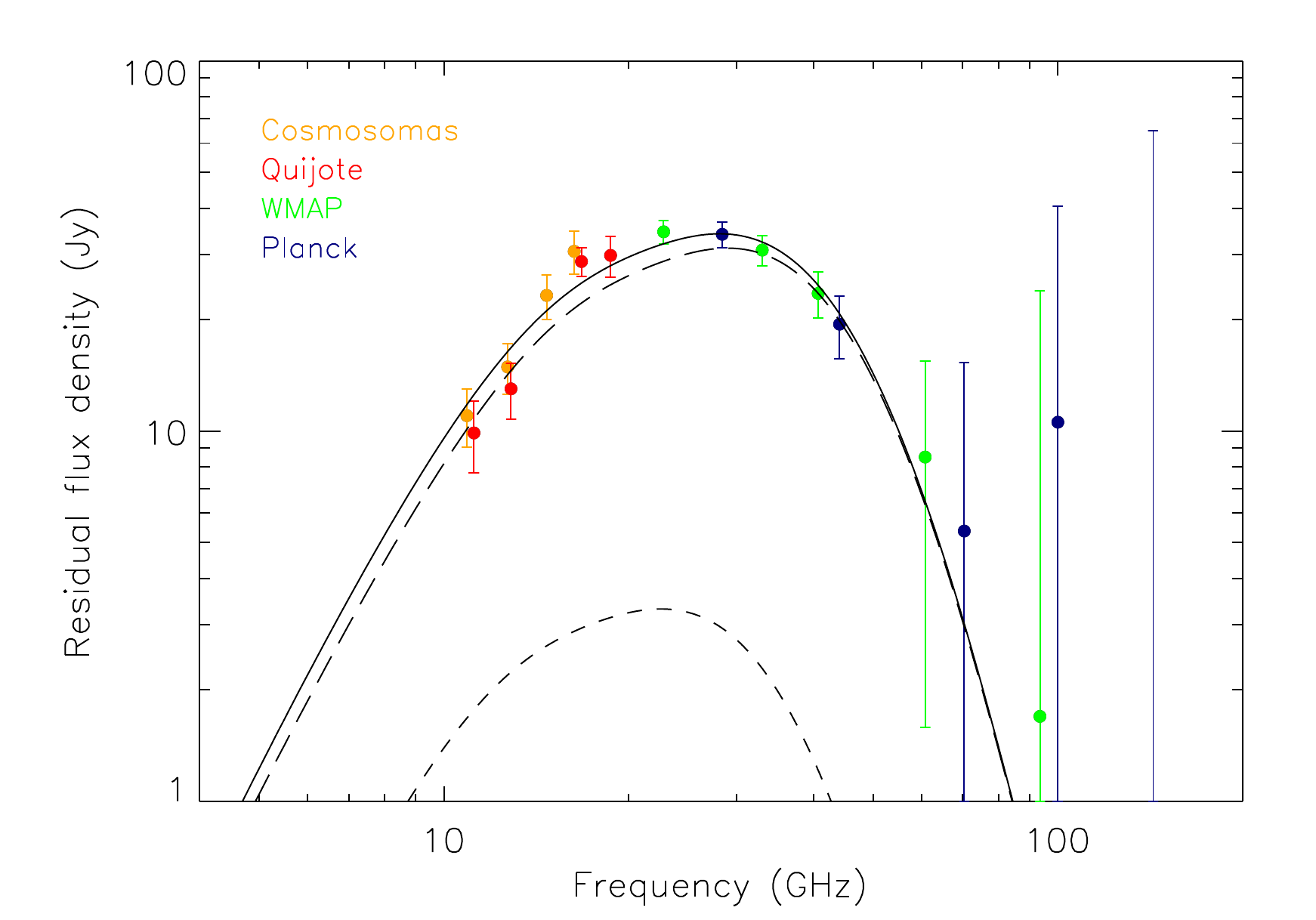}
\caption{\small Spectrum of the AME in G159.6-18.5 after subtracting the best-fit free-free, CMB and thermal dust components. The 
long-dashed line and the dashed line show respectively the spinning dust spectrum for the molecular and the atomic phases, while the 
solid line is the sum of the two.}
\label{fig:ressed}
\end{figure}

In Fig.~\ref{fig:ressed} we show the residual spinning dust spectrum, obtained after subtracting the best-fit free-free, CMB and thermal 
dust components.

\subsection{Polarization constraints}\label{sec:pol_limits}

As no clear emission is seen in the $Q$ and $U$ maps of Fig.~\ref{fig:perseus_maps}, we derive upper limits on the 
polarization fraction of G159.6-18.5, following the procedure explained in section~\ref{sec:methodology}. In 
section~\ref{sec:intensity_sed} we defined the sizes of the aperture and of the background annulus so that we reproduced the 
fluxes in \citet{per20}, which were obtained through Gaussian fitting. In order to minimize the error associated with the background 
subtraction, here we extend the size of the background annulus, and use a circular aperture with radius $1.5^\circ$, and a background 
ring between $1.5^\circ$ and $2.5^\circ$, with an extension towards the north-west as is shown in the maps of 
Fig.~\ref{fig:perseus_maps}. At each frequency we calculate the RMS levels in this background annulus, to define the 
quantities $\sigma(T_j)$ that are introduced in equation~\ref{eq:err2_flux} to calculate the errors on the Stokes parameters $Q$ and $U$. 
These errors, together with the fluxes resulting from the aperture photometry integration, are quoted in Table~\ref{tab:pol_limits}. Note 
that here the intensity fluxes in the QUIJOTE frequencies are slightly different to those presented in Table~\ref{tab:fluxes_sed} 
owing to the different apertures. In order to get the AME residual fluxes shown in Table~\ref{tab:pol_limits}, using the new fluxes, we repeat the same fit that was performed 
in section~\ref{sec:intensity_sed}, considering the same spinning dust spectra for the molecular and atomic phases.

It is important to point out that, as it became clear when we discussed the results of the 
jackknife tests in section~\ref{sec:maps}, the uncertainties on the $I$ fluxes are here 
biased high because some extended emission of the source leaks into the background annulus. This does not have significant 
implications in our analysis as the uncertainty in the polarization fraction is driven by the errors in $Q$ and $U$.

The $Q$ and $U$ fluxes shown in Table~\ref{tab:pol_limits} are consistent with zero, and therefore we obtain de-biased (in 
section~\ref{sec:methodology} we explained how this de-biasing is applied) upper limits at the 95\% confidence level on the 
polarized intensity, $P_{\rm db}$. We also show in Table~\ref{tab:pol_limits} upper limits at the 95\% confidence level on the 
polarization fraction, taking as reference both the total intensity ($\Pi_{\rm db}$) and the residual AME intensity 
($\Pi_{\rm AME,db}$). The values below the horizontal line in this table correspond to constraints obtained in maps that have been 
built by combining the two frequency bands of each horn. The most stringent upper limit we get on the polarization fraction is 
$\Pi_{\rm db}<2.85\%$, and is obtained after combining the maps at 16.7 and 18.7~GHz. Due to the decrease of the intensity flux at 
lower frequencies, the constraints at 11.2 and 12.9~GHz are less stringent.

Note that, under the reliable assumption that the free-free emission is unpolarized \citep{rybicki79}, any possible detection of polarization 
below $\nu \sim 30$~GHz, where the thermal dust is clearly sub-dominant, should in principle be ascribed to AME. One caveat to 
this hypothesis is the possible presence of a Faraday Screen (FS) hosting a strong and regular magnetic field, which could induce a 
rotation of the background polarized emission. This idea was proposed by \citet{reich09} to explain the high degree of 
polarization they detected towards G159.6-18.5 in observations from the Effelsberg telescope at 2.7~GHz. They suggest that this same 
mechanism could indeed be the responsible for the tentative polarized emission seen by \citet{battistelli06} at 11~GHz. Using as face 
value the rotation measure obtained by \citet{reich09}, RM$=190$~rad~m$^{-2}$, \citet{lopez11} estimated a polarization fraction of 
$\lesssim 0.2$\%, well below the upper limit at this frequency. Using this RM we estimate polarization fractions from the FS of 
$\approx 3.5\%$ and $\approx 0.4\%$, respectively at 11 and 19~GHz. These values are well below our upper limits, while the value 
at 11~GHz is compatible with the measurement of \citet{battistelli06}.

\begin{table*} 
\begin{center}
\begin{tabular}{ccccccccc}
\hline\hline
\noalign{\smallskip}
$\nu$ (GHz) & $I$ (Jy)  &  $I_{\rm AME}$ (Jy) & $Q$ (Jy) &  $U$ (Jy) &         $P_{\rm db}$ (Jy) & $\Pi_{\rm db}$ (\%) &$\Pi_{\rm AME,db}$ (\%) \\
\noalign{\smallskip}\hline\noalign{\smallskip} 
 11.2  &   14.0 $\pm$ 3.3   &	 9.4 $\pm$ 3.4  &    -0.05 $\pm$ 0.59	& -0.39 $\pm$ 0.47   &    $<1.19$  &  $<8.79$	& $<13.85$ \\
\noalign{\smallskip}
 12.9  &   17.5 $\pm$ 3.5   &	12.5 $\pm$ 3.5  &     0.45 $\pm$ 0.68	& -0.23 $\pm$ 0.56   &	  $<1.44$  &  $<8.46$	& $<12.20$ \\
\noalign{\smallskip}
 16.7  &   31.2 $\pm$ 3.0   &	28.2 $\pm$ 3.0  &     0.24 $\pm$ 0.47	& -0.83 $\pm$ 0.52   &	  $<0.95$  &  $<5.14$	& $<5.71$ \\
\noalign{\smallskip}
 18.7  &   31.5 $\pm$ 4.6   &	29.4 $\pm$ 4.7  &    -0.11 $\pm$ 0.75	&  0.18 $\pm$ 0.70   &	  $<1.49$  &  $<4.76$	& $<5.08$ \\ 
\noalign{\smallskip}\hline\noalign{\smallskip}
 12.0  &   15.5 $\pm$ 3.4   &   10.2 $\pm$ 3.4 	&     0.14 $\pm$ 0.38   & -0.36 $\pm$ 0.37   &    $<0.94$  &  $<6.26$   & $<10.10$ \\
\noalign{\smallskip}
 17.7  &   31.8 $\pm$ 3.7   &	26.7 $\pm$ 3.7 	&     0.06 $\pm$ 0.42   & -0.26 $\pm$ 0.40   &    $<0.91$  &  $<2.85$   & $<3.42$ \\
\noalign{\smallskip}
\hline\hline
\end{tabular}
\end{center}
\normalsize
\medskip
\caption[tab:pol_limits]{Intensity and polarization fluxes at the position of G159.6-18.5 determined through an aperture photometry 
integration, and upper limits on the polarized intensity and on the polarization fraction. We show the total intensity 
fluxes ($I$) as well as the residual AME fluxes ($I_{\rm AME}$) after subtracting the free-free and thermal dust components. Similarly, for the 
polarization fraction we show upper limits calculated with respect to the total and to the residual intensities. The error bars 
are at 1-sigma, whereas the upper limits are given at the 95\% confidence level. Results are shown for each of the four QUIJOTE 
frequency channels, as well as for a combination of the low and high frequency bands.}
\label{tab:pol_limits}
\end{table*}

In section~\ref{sec:maps} we mentioned that one important consistency test for our data processing is the verification that it 
shows no polarization at the position of un-polarized sources like the California HII region. In 
Table~\ref{tab:pol_limits_california} we show the corresponding $I$, $Q$ and $U$ fluxes, and derived upper limits on $P$ and $\Pi$, 
which have been obtained using the apertures shown in Fig.~\ref{fig:california_maps}. The limits on the polarization fraction 
stand at the level of $2$ to $4\%$ (95\% C.L.), depending on the frequency band. This constrains the possible existence of a 
leakage from intensity to polarization to values below this level. In fact, as it was mentioned in section~\ref{sec:quijote_data}, 
we expect the leakage to be below 0.3\% ($\leq -25$~dB). This number has been verified using observations of Cas A, where we recover 
a leakage pattern with a quadrupolar shape in the $Q$ and $U$ maps, with a level of around $0.2\%$ (further details will be presented 
in separate technical papers).

\begin{table*} 
\begin{center}
\begin{tabular}{ccccccc}
\hline\hline
\noalign{\smallskip}
$\nu$ (GHz) & $I$ (Jy)  & $Q$ (Jy) &  $U$ (Jy) &    $P_{\rm db}$ (Jy) & $\Pi_{\rm db}$ (\%) \\
\noalign{\smallskip}\hline\noalign{\smallskip} 
 $11.2$	&   $59.0\pm 4.1$ &  $-0.45\pm 0.36$	& $ 0.52\pm 0.42$   &	 $<1.25$  &   $<2.12$  \\
\noalign{\smallskip}
 $12.9$	&   $54.7\pm 4.2$ &  $-0.02\pm 0.44$	& $ 0.24\pm 0.46$   &	 $<0.96$  &   $<1.76$  \\
\noalign{\smallskip}
 $16.7$	&   $49.0\pm 3.7$ &  $-0.07\pm 0.38$	& $ 0.38\pm 0.35$   &	 $<0.91$  &   $<1.87$  \\
\noalign{\smallskip}
 $18.7$	&   $51.8\pm 6.2$ &  $-0.10\pm 0.52$	& $-0.68\pm 0.50$   &	 $<1.43$  &   $<2.82$  \\
\noalign{\smallskip}
\hline\hline
\end{tabular}
\end{center}
\normalsize
\medskip
\caption[tab:pol_limits_california]{Intensity and polarization fluxes of the California HII region determined through an aperture photometry 
integration, and upper limits on the polarized intensity and on the polarization fraction. The error bars of the $I$, $Q$ and $U$ fluxes 
are at 1-sigma, whereas the upper limits on the polarized intensity and on the polarization fraction are referred to the 95\% confidence level.}
\label{tab:pol_limits_california}
\end{table*}

\begin{figure}
\centering
\includegraphics[width=8.7cm]{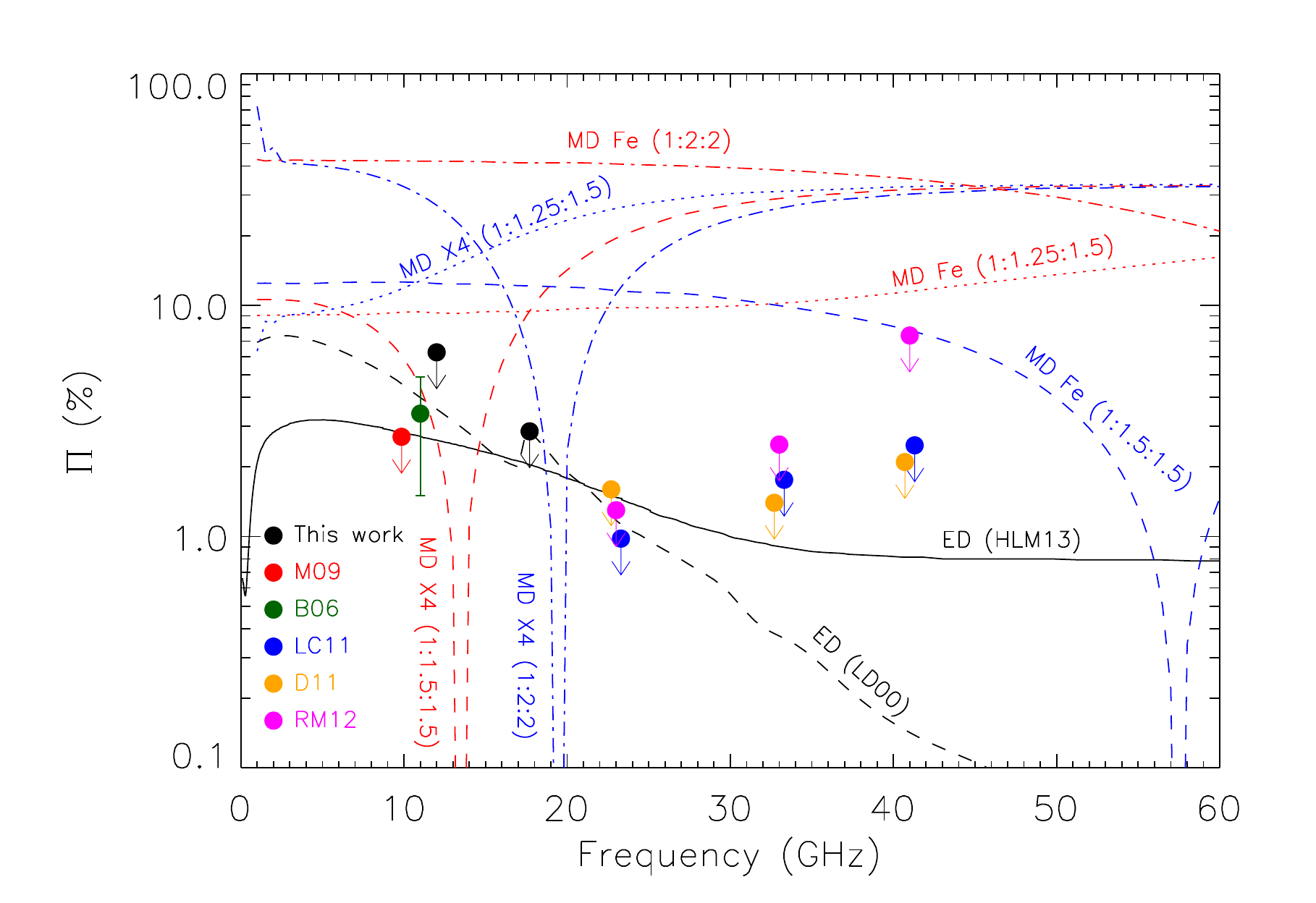}
\caption{\small Constraints (95\% C.L.) on several microwave emission models based on our upper limits on G159.6-18.5 (black dots) and 
on others in different regions from the literature 
\citep{mason09,battistelli06,lopez11,dickinson11,rubino12b}. The black lines are theoretical predictions for the frequency dependence of 
the fractional polarization of the electric dipole (ED) emission from \citet{lazarian00} (dashed line) and \citet{hoang13} (solid line). The red and blue 
lines correspond to the models of \citet{draine99} for the magnetic dipole (MD) emission. The different lines correspond to different grain compositions
 and geometries. The observations are generally consistent with the electric dipole emission models, and rule out the models for the 
 fractional polarization of the magnetic dipole emission.}
\label{fig:pol_constraints}
\end{figure}

We plot our constraints, and previous results in the literature at different frequencies, in Fig.~\ref{fig:pol_constraints}. It 
can be seen here that our observations fill the gap between previous results at frequencies below 11~GHz and above 20~GHz. 
The point at 9.65~GHz is an upper limit coming from GBT observations on LDN1622 \citep{mason09}. The value at 11~GHz represents 
the aforementioned tentative detection towards G159.6-18.5 at the $1.8\sigma$ level ($\Pi=3.4_{-1.9}^{+1.5}\%$) of AME polarization 
from the COSMOSOMAS experiment \citep{battistelli06}, which is still consistent with our upper bound at $12.0$~GHz. The points at $\nu>20$~GHz 
come from \wmap 7-year data on the Perseus \citep{lopez11} and $\rho$ Ophiuchi (Dickinson et al. 2011, they also present 
results on Perseus) molecular clouds and on the HII region [LPH96]~201.663+1.643 (Rubi\~no-Mart\'{\i}n et al. 2012b, they also show 
less-stringent constraints from the Pleiades reflection nebula and from the dark nebula LDN1622). 
Currently, the most stringent constraint is that obtained by \citet{lopez11}: $\Pi<0.98\%$ at 23~GHz. This limit benefits 
from the fact that it is obtained at a frequency close to the peak ($\approx 28$~GHz) of the AME in G159.6-18.5. The constraints 
on the polarization fraction from QUIJOTE come from a spectral region where the intensity flux density drops. However, according to the 
model of \citet{hoang13}, the polarization fraction of the spinning dust emission peaks at $\approx 5$~GHz, and therefore 
observations at lower frequencies are more appropriate to constrain this model. All measurements, including ours, are 
consistent with this model, except maybe the previous limit from \citet{lopez11} and that from \citet{mason09}, which are 
slightly below. 

Our upper limits on $\Pi$ are not only affected by the decrease of AME flux at our frequencies, but also by the sparse sky coverage 
of our observations. We decided to survey an area large enough to cover the California HII region and also to get the source 
simultaneously in the four horns. Currently we are undertaking observations on a smaller area, centred in two individual horns, with 
the goal to increase by a factor $8$ the integration time per unit area. This would improve our map sensitivity by a factor 
$2.8$, which would help push our current upper limits down to $\approx 1.7\%$ at $18.7$~GHz, and $\approx 1.0\%$ after combining the 
$16.7$ and $18.7$~GHz frequency bands, which is well below the polarization degree predicted by the \citet{hoang13} model at these 
frequencies. Furthermore, future observations at 30~GHz with the upcoming TGI experiment, which is expected to be 13 times more 
sensitive than the current MFI, will push these upper limits down by a significant amount.

However, it is important to note that the curve of the polarization degree obtained by \citet{hoang13} represents 
an upper limit. They inferred the alignment efficiency of interstellar dust grains from observations of the UV polarization 
excess towards two stars to predict the polarization degree of the spinning dust emission. In so doing they assumed that the UV 
polarization bump is produced exclusively by polycyclic aromatic hydrocarbon (PAHs) molecules, those which are thought to be 
responsible for spinning dust emission. However, if the graphite grains were also aligned, contributing to the UV 
polarization bump, then the alignment efficiency of PAHs could actually be lower and so would be the inferred degree of spinning 
dust polarization.

The previous model from \citet{lazarian00} for spinning dust polarization is also plotted in Fig.~\ref{fig:pol_constraints}, 
for the case of a cold neutral medium (CNM), together with different models from \citet{draine99} predicting the degree of 
polarization of the magnetic dipole emission for different grain shapes and compositions. In particular, we show the cases of 
grains made of metallic Fe, and of the hypothetical material X4 defined in \citet{draine99}. Concerning the grain geometries, 
different axial ratios $a1$:$a2$:$a3$ are shown, as indicated in the figure. All these models predict polarization fractions 
typically above $10\%$, much higher than that of the spinning dust emission, and are ruled out by our and previous observations. 
However, it must be noted that these models consider the dust grains to have a perfect ordering of the atomic magnetic moments in 
a single domain. A more realistic case considering randomly-oriented single-domain magnetic inclusions has been recently studied by 
\citet{draine13}, resulting in much lower polarization degrees, of the order of $5\%$ in the range $10-20$~GHz, or below.

\section{Conclusions}

We have presented the first results obtained with the QUIJOTE experiment, a new polarimeter aimed at measuring the B-mode anisotropy 
from inflation and also at characterizing the polarized foregrounds at low frequencies. These results are based on observations of 
the Perseus molecular complex, one of the regions where the AME has extensively been studied. Our observations cover G159.6-18.5, 
the region where AME is produced, and the California HII region. Our intensity data on G159.6-18.5 traces the decrease of the flux of 
this source at frequencies below $\sim 20$~GHz, confirming the prediction of the models based on spinning dust radiation, which 
currently are the best at reproducing the frequency spectrum of the AME. This confirms previous results on this region obtained 
with the COSMOSOMAS experiment \citep{watson05}. When combining the QUIJOTE measurements with data coming from COSMOSOMAS, 
{\it WMAP}-9yr and {\it Planck}-DR1, we get an intensity SED with 13 points between 10 and 50~GHz being dominated by AME, providing what probably 
is the most precise spectrum of this emission ever measured in an individual object. This intensity SED is well fitted 
($\chi^2_{\rm red}=0.99$) by a combination of free-free, CMB, spinning dust and thermal dust, and two spinning dust components associated 
to a high-density molecular phase and to a low-density atomic phase. We attempt to fit some of the parameters describing the physical 
environment of the cloud, and which define the shape of the spinning dust spectrum, but the solution is clearly degenerate and 
this approach does not achieve a better fit. However, with the physical parameters we use for the spinning dust spectra, we get 
reliable values for the 7 parameters of our model.

The California HII region is well detected in intensity, presenting a spectrum at QUIJOTE frequencies consistent with free-free 
emission. The polarization maps at the position of California are consistent with noise, as it is expected for regions dominated by 
free-free emission, which is very lowly polarized. No polarization is detected at the position of G159.6-18.5 either, so then we 
extract upper limits on the polarized intensity and on its degree of polarization. After combining the low-frequency and 
high-frequency maps of two horns of QUIJOTE, we get $\Pi<6.26\%$ and $\Pi<2.85\%$ ($95\%$ C.L.), respectively at $12.0$ and 
$17.7$~GHz, being the first upper limits on the AME polarization fraction in this frequency range. These constraints are consistent 
with models of the spinning dust emission, that predict polarization fractions between $2$ and $3\%$ in this frequency range, and 
rule out several models based on magnetic dipole emission. Pushing these constraints even further is crucial to test the 
theoretical models and also to assess up to what level AME will hinder the detection of the B-mode signal by current and future 
experiments operating in the frequency range 10 to 80~GHz. New QUIJOTE observations in this region will improve the sensitivity 
in our maps so that we will be able to improve the previous constraints at least by a factor 2, allowing to reach a sensitivity 
on the polarization fraction close to the predictions of the spinning dust models.

\section*{Acknowledgments}
This work has been partially funded by the Spanish Ministry of of Economy and
Competitiveness (MINECO) under the projects AYA2007-68058-C03-01, AYA2010-21766-C03-02, and the Consolider-Ingenio project
CSD2010-00064 (EPI: Exploring the Physics of Inflation). 
CD acknowledges support from an ERC Starting (Consolidator) Grant (no.~307209). SH acknowledges support from an 
STFC-funded studentship.
We acknowledge the use of data from the Planck/ESA mission, downloaded from the Planck Legacy Archive, and of the 
Legacy Archive for Microwave Background Data Analysis (LAMBDA). Support for LAMBDA is provided 
by the NASA Office of Space Science. Some of the results in this paper have been derived using the HEALP{\sc ix} \citep{gorski05} 
package.

\bibliographystyle{mn2e}

\begin{thebibliography}{}

\bibitem[Ali-Ha{\"i}moud et al.(2009)]{ali09} Ali-Ha{\"i}moud, Y., Hirata, C.~M.,
\& Dickinson, C.\ 2009, \mnras, 395, 1055

\bibitem[AMI Consortium et al.(2009)]{ami09} AMI Consortium, 
Scaife, A.~M.~M., Hurley-Walker, N., et al.\ 2009, \mnras, 400, 1394 

\bibitem[Battistelli et al.(2006)]{battistelli06} Battistelli, E.~S., 
Rebolo, R., Rubi{\~n}o-Mart{\'{\i}}n, J.~A., et al.\ 2006, \apjl, 645, L141 

\bibitem[Bennett et al.(2013)]{bennett13} Bennett, C.~L., Larson, 
D., Weiland, J.~L., et al.\ 2013, \apjs, 208, 20 

\bibitem[Berkhuijsen(1972)]{berkhuijsen72} Berkhuijsen, E.~M.\ 1972, \aaps, 5, 263

\bibitem[BICEP1 Collaboration et al.(2014)]{bicep114} BICEP1 Collaboration, 
Barkats, D., Aikin, R., Bischoff, C., et al.\ 2014, \apj, 783, 67

\bibitem[BICEP2 Collaboration et al.(2014)]{bicep214} BICEP2 Collaboration, Ade, P.~A.~R., Aikin, 
R.~W., Barkats, D., et al.\ 2014, Physical Review Letters, 112, 241101 

\bibitem[Casassus et al.(2006)]{casassus06} Casassus, S., Cabrera,
G.~F., F{\"o}rster, et al.\ 2006, \apj, 639, 951

\bibitem[Casassus et al.(2008)]{casassus08} Casassus, S., Dickinson, C., Cleary, K.,
et al.\ 2008, \mnras, 391, 1075

\bibitem[Cernicharo et al.(1985)]{cernicharo85} Cernicharo, J., Bachiller, R., \& Duvert, G.\ 1985, \aap, 149, 273 

\bibitem[Davies et al.(2006)]{davies06} Davies, R.~D., Dickinson, C., Banday, A.~J., et al.\ 2006, \mnras, 370, 1125 

\bibitem[Dickinson et al.(2009)]{dickinson09} Dickinson, C., Davies, R.~D., Allison, J.~R., et al.\ 2009, \apj, 690, 1585 

\bibitem[Dickinson et al.(2011)]{dickinson11} Dickinson, C., Peel, 
M., \& Vidal, M.\ 2011, \mnras, 418, L35

\bibitem[Draine \& Lazarian(1998)]{draine98} Draine, B.~T., \& Lazarian, A.\ 1998, \apj, 508, 157 

\bibitem[Draine \& Lazarian(1999)]{draine99} Draine, B.~T., \& Lazarian, A.\ 1999, \apj, 512, 740 

\bibitem[Draine \& Hensley(2013)]{draine13} Draine, B.~T., \& Hensley, B.\ 2013, \apj, 765, 159 

\bibitem[Femenia et al.(1998)]{femenia98} Femenia, B., Rebolo, 
R., Gutierrez, C.~M., Limon, M., \& Piccirillo, L.\ 1998, \apj, 498, 117 

\bibitem[Finkbeiner et al.(2002)]{finkbeiner02} Finkbeiner, D.~P.,
Schlegel, D.~J., Frank, C., \& Heiles, C.\ 2002, \apj, 566, 898

\bibitem[Finkbeiner et al.(2004)]{finkbeiner04} Finkbeiner, D.~P., 
Langston, G.~I., \& Minter, A.~H.\ 2004, \apj, 617, 350 

\bibitem[Flauger et al.(2014)]{flauger14} Flauger, R., Hill, 
J.~C., \& Spergel, D.~N.\ 2014, arXiv:1405.7351 

\bibitem[Garc{\'{\i}}a-Lorenzo et al.(2010)]{garcia10} 
Garc{\'{\i}}a-Lorenzo, B., Eff-Darwich, A., Castro-Almaz{\'a}n, J., et al.\ 
2010, \mnras, 405, 2683 

\bibitem[G{\'e}nova-Santos et al.(2011)]{genova11} 
G{\'e}nova-Santos, R., Rebolo, R., Rubi{\~n}o-Mart{\'{\i}}n, J.~A., 
L{\'o}pez-Caraballo, C.~H., \& Hildebrandt, S.~R.\ 2011, \apj, 743, 67 

\bibitem[G{\'o}rski et al.(2005)]{gorski05} G{\'o}rski, K.~M., 
Hivon, E., Banday, A.~J., et al.\ 2005, \apj, 622, 759 

\bibitem[Hafez et al.(2008)]{hafez2008} Hafez, Y.~A., Davies, 
R.~D., Davis, R.~J., et al.\ 2008, \mnras, 388, 1775 

\bibitem[Hauser et al.(1998)]{hauser98} Hauser, M.~G., Arendt, R.~G., Kelsall, T., et al.\ 1998, \apj, 508, 25

\bibitem[Haslam et al.(1982)]{haslam82} Haslam, C.~G.~T., Salter, C.~J., Stoffel, H., \& Wilson, W.~E.\ 1982, \aaps, 47, 1

\bibitem[Hoang et al.(2010)]{hoang10} Hoang, T., Draine, B.~T., \& Lazarian, A.\ 2010, \apj, 715, 1462

\bibitem[Hoang et al.(2013)]{hoang13} Hoang, T., Lazarian, A., 
\& Martin, P.~G.\ 2013, \apj, 779, 152 

\bibitem[Kamionkowski et al.(1997)]{kamionkowsky97} Kamionkowski, M., 
Kosowsky, A., \& Stebbins, A.\ 1997, \prd, 55, 7368 

\bibitem[Kogut et al.(1996a)]{kogut96a} Kogut, A., Banday, A.~J.,
Bennett, C.~L., et al.\ 1996a, \apj, 460, 1

\bibitem[Kogut et al.(1996b)]{kogut96b} Kogut, A., Banday, A.~J.,
Bennett, C.~L., et al.\ 1996b, \apjl, 464, L5

\bibitem[Lazarian 
\& Draine(2000)]{lazarian00} Lazarian, A., \& Draine, B.~T.\ 2000, \apjl, 536, L15 

\bibitem[Leitch et al.(1997)]{leitch97} Leitch, E.~M., Readhead,
A.~C.~S., Pearson, T.~J., \& Myers, S.~T.\ 1997, \apjl, 486, L23

\bibitem[L{\'o}pez-Caraballo et al.(2011)]{lopez11} 
L{\'o}pez-Caraballo, C.~H., Rubi{\~n}o-Mart{\'{\i}}n, J.~A., Rebolo, R., 
\& G{\'e}nova-Santos, R.\ 2011, \apj, 729, 25 

\bibitem[Macellari et al.(2011)]{macellari11} Macellari, N., Pierpaoli, E., Dickinson, C., 
\& Vaillancourt, J.~E. \ 2011, \mnras, 418, 888

\bibitem[Mason et al.(2009)]{mason09} Mason, B.~S., Robishaw, 
T., Heiles, C., Finkbeiner, D., \& Dickinson, C.\ 2009, \apj, 697, 1187 

\bibitem[Mather et al.(1999)]{mather99} Mather, J.~C., Fixsen,
D.~J., Shafer, R.~A., Mosier, C., \& Wilkinson, D.~T.\ 1999, \apj, 512, 511

\bibitem[Mortonson \& Seljak(2014)]{mortonson14} Mortonson, M.~J., \& Seljak, U.\ 2014, arXiv:1405.5857 

\bibitem[Murphy et al.(2010)]{murphy10} Murphy, E.~J., Helou, G., Condon, J.~J., et al.\ 2010, \apjl, 709, L108 

\bibitem[Planck Collaboration et al.(2011)]{per20} Planck Collaboration.
Planck Early Results XX, 2011, \aap, 536, A20 

\bibitem[Planck Collaboration et al.(2014a)]{pip25} Planck Collaboration.
Planck Intermediate Results XXV, 2014a, \aap, 565, AA103 

\bibitem[Planck Collaboration et al.(2014b)]{pip30} Planck Collaboration.
Planck Intermediate Results XXX, 2014b, arXiv:1409.5738 

\bibitem[Planck Collaboration et al.(2014c)]{cpp1} Planck Collaboration.
Planck 2013 Resuls I, 2014c, \aap, 571, AA1

\bibitem[Planck Collaboration et al.(2014d)]{cpp12} Planck Collaboration.
Planck 2013 Resuls XII, 2014d, \aap, 571, AA12

\bibitem[Planck Collaboration et al.(2014e)]{cpp13} Planck Collaboration.
Planck 2013 Resuls XIII, 2014e, \aap, 571, AA13

\bibitem[QUIET Collaboration et al.(2012)]{quiet12} QUIET 
Collaboration, Araujo, D., Bischoff, C., et al.\ 2012, \apj, 760, 145 

\bibitem[de Oliveira-Costa et al.(1998)]{oliveira98} de
Oliveira-Costa, A., Tegmark, M., Page, L.~A.,
\& Boughn, S.~P.\ 1998, \apjl, 509, L9

\bibitem[de Oliveira-Costa et al.(1999)]{oliveira99} de
Oliveira-Costa, A., Tegmark, M., Gutierrez, C.~M., et al.\ 1999, \apjl, 527, L9

\bibitem[Platania et al.(2003)]{platania03} Platania, P., Burigana, C., Maino, D., et al.\ 2003, \aap, 410, 847

\bibitem[Rebolo et al., in prep.]{rebolo_prep} Rebolo, R., et al.\, in preparation

\bibitem[Reich \& Reich(1986)]{reich86} Reich, P., \& Reich, W.\ 1986, \aaps, 63, 205

\bibitem[Reich \& Reich(2009)]{reich09} Reich, W., \& Reich, P.\ 2009, IAU Symposium, 259, 603 

\bibitem[Rybicki \& Lightman(1979)]{rybicki79} Rybicki, G.~B., \& Lightman, A.~P.\ 1979, 
{\it Radiative Processes in Astrophysics}, John Wiley \& Sons

\bibitem[Rubi{\~n}o-Mart{\'{\i}}n et al.(2012a)]{rubino12} 
Rubi{\~n}o-Mart{\'{\i}}n, J.~A., Rebolo, R., Aguiar, M., et al.\ 2012a, 
{\it Society of Photo-Optical Instrumentation Engineers (SPIE) Conference Series}, 8444

\bibitem[Rubi{\~n}o-Mart{\'{\i}}n et al.(2012b)]{rubino12b} 
Rubi{\~n}o-Mart{\'{\i}}n, J.~A., L\'opez-Caraballo, C.~H., G\'enova-Santos, R., \& Rebolo, R. 
\ 2012b, Advances in Astronomy, 351836

\bibitem[Silsbee et al.(2011)]{silsbee11} Silsbee, K., Ali-Ha{\"i}moud, Y., \& Hirata, C.~M.\ 2011, \mnras, 411, 2750

\bibitem[Tibbs et al.(2010)]{tibbs10} Tibbs, C.~T., Watson, R.~A., Dickinson, C., et al.\
2010, \mnras, 402, 1969

\bibitem[Tibbs et al.(2013)]{tibbs13} Tibbs, C.~T., Scaife, 
A.~M.~M., Dickinson, C., et al.\ 2013, \apj, 768, 98 

\bibitem[Vaillancourt(2006)]{vaillancourt06} Vaillancourt, J.~E.\ 2006, \pasp, 118, 1340 

\bibitem[Vidal et al.(2011)]{vidal11} Vidal, M., Casassus, S., Dickinson, C., et al.\ 2011, \mnras, 414, 2424 

\bibitem[Watson et al.(2005)]{watson05} Watson, R.~A., Rebolo,
R., Rubi{\~n}o-Mart{\'{\i}}n, et al.\ 2005, \apjl, 624, L89

\bibitem[Weiland et al.(2011)]{weiland11} Weiland, J.~L., 
Odegard, N., Hill, R.~S., et al.\ 2011, \apjs, 192, 19

\bibitem[Weingartner \& Draine(2001)]{weingartner01} Weingartner, J.~C., \& Draine, B.~T.\ 2001, \apj, 548, 296

\bibitem[Ysard \& Verstraete(2010)]{ysard10} Ysard, N., \& Verstraete, L.\ 2010, \aap, 509, AA12 

\bibitem[Zaldarriaga \& Seljak(1997)]{zaldarriaga97} Zaldarriaga, M., \& Seljak, U.\ 1997, \prd, 55, 1830 

\end{thebibliography}

\pagestyle{plain}

\bsp
\label{lastpage}
\end{document}